\pdfoutput=1
\documentclass[11pt]{article}

\usepackage[]{ACL2023}

\usepackage{times}
\usepackage{latexsym}
\usepackage{booktabs}
\usepackage{graphicx}
\usepackage{color}
\usepackage{xcolor}
\usepackage{bbding}
\usepackage{pifont}
\usepackage{multirow}
\usepackage[T1]{fontenc}
\usepackage[utf8]{inputenc}

\usepackage{microtype}

\usepackage{inconsolata}
\usepackage{pifont} 

\title{Multimodal Recommendation Dialog with Subjective Preference: \\ A New Challenge and Benchmark}

\author {
    Yuxing Long\textsuperscript{\rm1},
    Binyuan Hui\textsuperscript{\rm2},
    Caixia Yuan\textsuperscript{\rm1},
    Fei Huang\textsuperscript{\rm2},
    Yongbin Li\textsuperscript{\rm2 \footnotemark[1]},
    Xiaojie Wang\textsuperscript{\rm1 \footnotemark[1]} \\
    \textsuperscript{\rm 1} Beijing University of Posts and Telecommunications, 
    \textsuperscript{\rm 2} Independent Researcher \\
    \texttt{\{longyuxing,yuancx,xjwang\}@bupt.edu.cn, lyb821@gmail.com}
}

\begin{document}
\maketitle
\renewcommand{\thefootnote}{\fnsymbol{footnote}}
\footnotetext[1]{Corresponding authors.}
\footnotetext[2]{The dataset and the code of the baseline model are available at https://github.com/LYX0501/SURE.}
\begin{abstract}
Existing multimodal task-oriented dialog data fails to demonstrate the diverse expressions of user subjective preferences and recommendation acts in the real-life shopping scenario. This paper introduces a new dataset SURE (\textit{Multimodal Recommendation Dialog with \textbf{SU}bjective P\textbf{RE}ference}), which contains 12$K$ shopping dialogs in complex store scenes. The data is built in two phases with human annotations to ensure quality and diversity. SURE is well-annotated with subjective preferences and recommendation acts proposed by sales experts. A comprehensive analysis is given to reveal the distinguishing features of SURE. Three benchmark tasks are then proposed on the data to evaluate the capability of multimodal recommendation agents. Based on the SURE, we propose a baseline model\footnotemark[2], powered by a state-of-the-art multimodal model, for these tasks. 
\end{abstract}

\section{Introduction}
Building conversational agents that can communicate with people in multimodal situations is an attractive goal for the AI community. Many different tasks and datasets for the multimodal dialog have been proposed in recent years. Among them, Moon et al.~\cite{simmc1} provided a multimodal task-oriented dialog dataset SIMMC 1.0 in two shopping domains. It aims to train interactive assistants which can handle multimodal inputs in a co-observed environment. The SIMMC challenge based on SIMMC 1.0 was held as part of DSTC9~\cite{DSTC9}. 
Many works~\cite{simmc1kung, simmc1sogang1, simmc1sogang2, simmc1astar, simmc1sense} have been done following SIMMC 1.0. Since SIMMC 1.0 environment is simple and far from realistic stores, Kottur et al~\cite{simmc2} proposed SIMMC 2.0 with closer-to-real-world shopping scenarios, which was then used in DSTC10 challenge ~\cite{DSTC10}. 

Though these datasets facilitate research of conversational agents, they simplify some crucial problems in the real-life shopping dialog, which should be addressed for building multimodal recommendation agents.

\begin{figure}[t]
    \centering
    \includegraphics[width=0.9\linewidth]{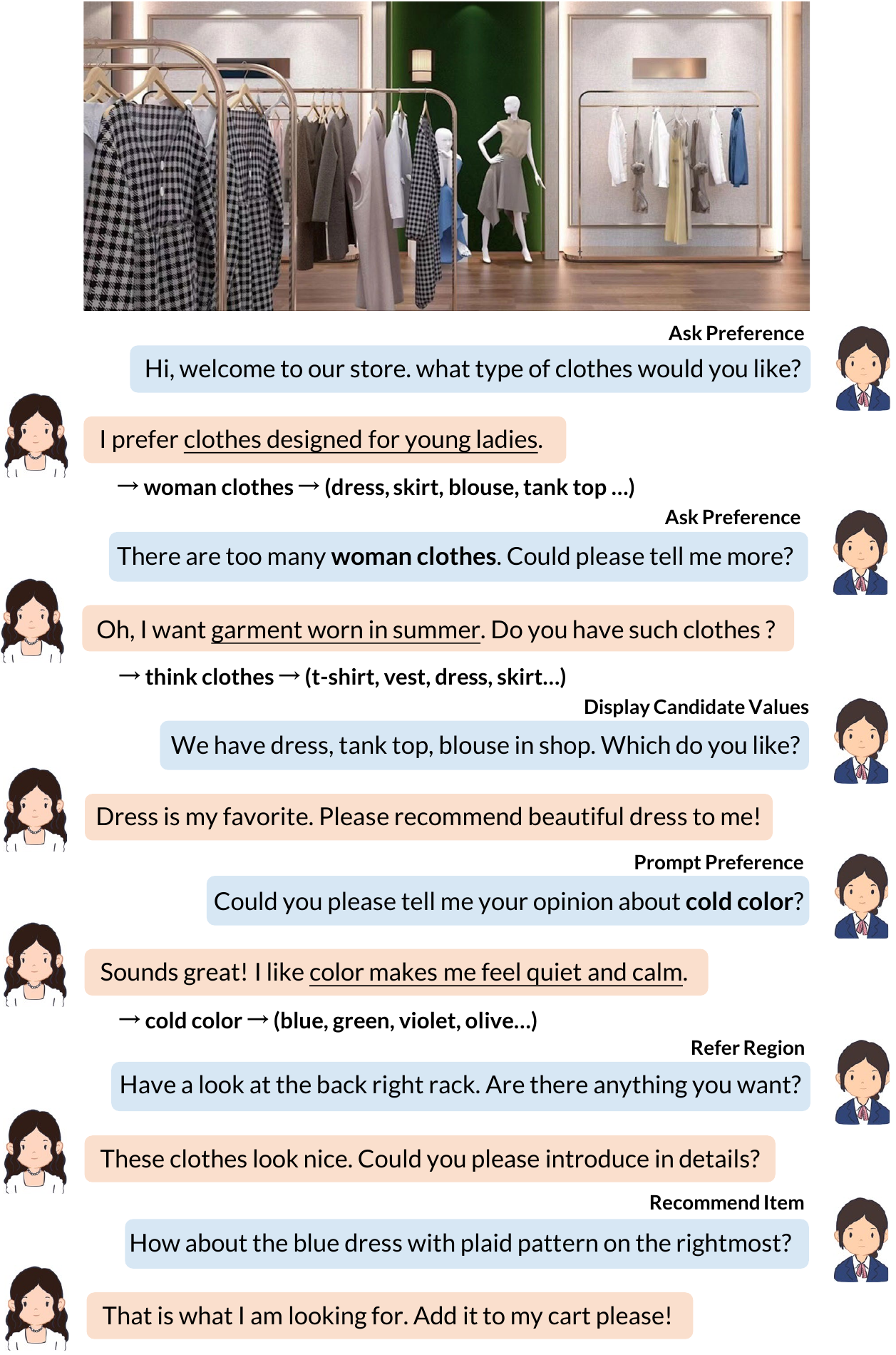}
    \caption{\textbf{Customer} \textit{(left)} expresses subjective preferences. \textbf{Salesperson} \textit{(right)} adopts different acts to clarify subjective preferences by multimodal context and recommend clothes from the situated store. Subjective preferences are highlighted by underlining while corresponding categorization concepts and candidate attribute values of scene items are shown in beneath.}
    \vspace{-0.5cm}
    \label{fig:introduction}
\end{figure}

In previous TOD datasets, most of the user descriptions for items (e.g.,~clothes or furniture) are referring expressions in the domains, such as \textit{"the white couch chair"}, and \textit{"the black hat in the middle of the long rack"}, which can be mapped to a slot value without ambiguity. While in practice, customers are not experts in the domains. They use lots of different words to describe what they want, such as \textit{"clothes designed for young ladies"}, and \textit{"color makes me feel quiet and calm"} in Fig.~\ref{fig:introduction}. These words (or phrases) usually reflect customers' subjective cognition and preference for items they want. We call this kind of expression as \textbf{subjective preference}. To understand such expressions, the salesperson needs to map subjective preferences to standardized categorization concepts in the domains, and then use the concepts to filter candidate attribute values (slot values) of scene items. (Fig.~\ref{fig:introduction} shows the two-step mappings under the subjective preferences). As we can see, such subjective preferences often correspond to a set of slot values instead of a unique one, which is very different from that in previous datasets. Facing this kind of customer requirement, a salesperson needs to communicate with the customers, utilize suitable strategies to narrow candidates progressively and give sound recommendations through multimodal context (Fig.~\ref{fig:introduction} shows some recommendation acts, such as \textit{Ask Preference}). In a word, understanding subjective preferences, finding a way to clarify the subjective preferences, and finally giving good recommendations is the challenge not depicted in both SIMMC 1.0 and 2.0. None of the existing multimodal dialogs research focuses on subjective preference and item recommendation.

\iffalse
\begin{table*}
    \centering
    \scalebox{0.6}{
    \begin{tabular}{lccccc}
        \toprule
         {\textbf{\textsc{Dataset}}} & \textbf{Domain} & \textbf{\# Sessions} & \textbf{Multimodal} & \textbf{Recommendation Strategy} & \textbf{Subjective Preference}   \\
        \midrule
         ReDial~\cite{li2018conversational}                   &Movie  &10k &\XSolidBrush   &\XSolidBrush   &\XSolidBrush    \\
         TG-ReDial~\cite{zhou2020topicguided}                 &Movie  &10k &\XSolidBrush   &\Checkmark   &\XSolidBrush     \\
         GoRecDial~\cite{DBLP:journals/corr/abs-1909-03922}   &Movie  &9k &\XSolidBrush   &\Checkmark   &\XSolidBrush    \\
         INSPIRED~\cite{inspired}                             &Movie  &1k  &\XSolidBrush   &\Checkmark   &\XSolidBrush   \\
         OpenDialKG~\cite{opendialkg}                         &Movie, Music, Book, Sport  &3k &\XSolidBrush  &\Checkmark   &\XSolidBrush    \\
         MMD~\cite{mmd}               &Fashion  &130.6k &\Checkmark   &\XSolidBrush  &\XSolidBrush   \\
         SIMMC 2.0~\cite{simmc2}      &Fashion, Furniture  &11.2k  &\Checkmark   &\XSolidBrush  &\XSolidBrush    \\
         \midrule
         SURE                      &Fashion, Furniture  &12k  &\Checkmark   &\Checkmark   &\Checkmark    \\
        \bottomrule
    \end{tabular}}
    \caption{Comparison of related conversational recommendation datasets and multimodal shopping datasets.}
    \vspace{-0.5cm}
    \label{tab:comparision}
\end{table*}
\fi

To facilitate building conversational agents that can handle subjective preference and make shopping recommendations, we introduce a dataset for \textit{Multimodal Recommendation Dialog with \textbf{SU}bjective P\textbf{RE}ference} (SURE). Specifically, we collect 12K salesperson $\leftrightarrow$ customer goal-oriented recommendation dialogs in complex store scenes in two phases. Dialog flows were first generated by self-playing between the carefully designed customer and salesperson simulators, which helps to ensure the flows are reasonable. Crowd-sourcing is then employed for rewriting categorization concepts in dialog flows to diverse subjective preferences. The dataset contains well-annotated subjective preferences and diverse dialog acts proposed by experienced sales experts, which provides rich resources for evaluating subjective preferences understanding and dialog policies.

We then propose three benchmark tasks for evaluating multimodal recommendation agents' capability on subjective preference understanding and item recommendation: \textbf{Subjective Preference Disambiguation}, \textbf{Referred Region Understanding}, and \textbf{Multimodal Recommendation}. We provide a baseline model for these tasks and highlight the key challenges and future research directions.

Our main contributions are:
\begin{itemize}
    \item[$\bullet$] We introduce a large-scale multimodal dialogs data in two domains including 12$K$ dialogs in complex scenes. The data is built in two phases with human annotations to ensure both dialog quality and language diversity. 
    \item[$\bullet$] The data is well-annotated with subjective preferences and recommendation acts. Diverse acts and transition probabilities are obtained from the survey for sales experts. 
    \item[$\bullet$] Three tasks are designed to evaluate the capability of multimodal recommendation agents. A strong baseline model MRA is proposed for these tasks.
\end{itemize}

\begin{figure*}[htp]
    \centering
    \includegraphics[width=0.9\linewidth]{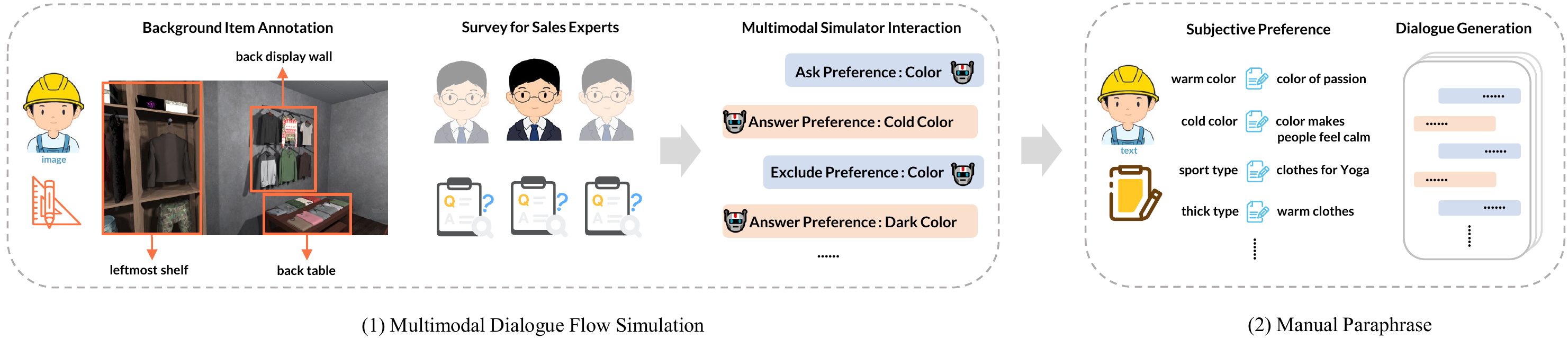}
    \caption{The annotation process of SURE dataset. (1)\,\textbf{Multimodal Dialog Flow Simulation}: we first hire annotators to label background items in the scene and do a survey for sales experts about attribute categorization and act. Then, we generate dialog flows by simulator interaction based on survey results. (2)\,\textbf{Manual Paraphrase}: annotators write subjective preferences based on attribute categorization concepts and paraphrase dialog flows.}
    \vspace{-0.5cm}
    \label{fig:annotation}
\end{figure*}

\section{Related Work}
\paragraph{Textual Conversational Recommendation.} 
The rise of e-books, music, and video websites has witnessed the development of conversational recommendation agents. Existing agents all operate in the textual modal, which elicits user preferences via conversation and recommends items based on dialog history and structured attribute data. The ReDial~\cite{li2018conversational} agent immediately recommends movies obtained from DBpedia~\cite{DBLP:conf/semweb/AuerBKLCI07} after the user expresses their preferences. TG-ReDial~\cite{zhou2020topicguided} is developed by walking along ConceptNet~\cite{DBLP:journals/corr/SpeerCH16} threads containing movies to collect user preferences and movie recommendations. GoRecDial~\cite{DBLP:journals/corr/abs-1909-03922} dialogs are collected by game-play to recommend target movies from candidate sets. DuRecDial~\cite{DBLP:journals/corr/abs-2005-03954} include Chinese dialogs between movie seeker and conversational bot based on the knowledge graph. INSPIRED~\cite{hayati-etal-2020-inspired} focuses on how social strategies adopted by the agent influence the final success rate of recommendation. Although these works study conversational recommendation problems from different aspects including topic, strategy, and language, they are all based on textual modal. Therefore, conversational agents built on these datasets cannot equip the abilities required in multimodal scenarios. Besides, all of these researches ignore developing agents to respond to subjective preferences frequently appearing in real recommendation dialogs.

\paragraph{Multimodal Shopping Dialogs.} 
Developing multimodal conversational agent for shopping scenarios is significant for improving the quality of commercial service quality. MMD~\cite{mmd} establishes the first large-scale multimodal dialog dataset between shoppers and sales agents, which empowers conversational agents with abilities of multimodal understanding and querying. SIMMC~\cite{simmc1} serves as a first step towards building task-oriented multimodal conversational agents with simple acts like informing information, confirming, and prompting. SIMMC 2.0~\cite{simmc2} constructs more complex multimodal context with closer-to-real-world store scenes and introduces new challenges like multimodal co-reference resolution and multimodal dialog state tracking. The SIMMC challenge based on SIMMC 1.0 was held as part of DSTC9~\cite{DSTC9}, there are many follows-up work~\cite{simmc1kung, simmc1sogang1, simmc1sogang2, simmc1astar, simmc1sense, spring}. SIMMC 2.0 was used in the DSTC10 challenge ~\cite{DSTC10}. These datasets facilitate research on multimodal conversational agents greatly. However, they still lack diverse expressions of user subjective preferences and recommendation acts.

\begin{table*}
    \centering
    \scalebox{0.54}{
    \begin{tabular}{cll}
        \toprule
         {\textbf{\textsc{Attribute Type}}} & \textbf{\textsc{Categorization Concept}} &  \textbf{\textsc{Subjective Preference}}  \\
        \midrule
        \multirow{5}*{\textbf{Color}} 
        & warm color (\textit{red, brown, yellow, light pink...})  
        & \textit{"color of passion”, "lively color", "color for outgoing people", "color for happy activity"...} \\ \cmidrule(r){2-3}
        & cold color (\textit{green, blue, light purple, olive...}) 
        & \textit{"cool color", "color which makes people feel calm", "refreshing color", “color of quietness”...} \\ \cmidrule(r){2-3}
        & powerful color (\textit{red, orange, light red...}) 
        & \textit{"color full of energy", "color inspires the fighting spirit", "color makes people feel excited"...} \\ \cmidrule(r){2-3}
        & mysterious color (\textit{violet, black, dark blue...}) 
        & \textit{"elusive color", "unfathomable color", "color inspires exploration", "color arouses curiosity"...} \\ \midrule
        \multirow{4}*{\textbf{Pattern}} 
        & lively pattern (\textit{floral pattern, leopard print...})  
        & \textit{"vibrant pattern", "pattern closer to nature", "pattern that is popular among conservationists"...} \\ \cmidrule(r){2-3}
        & dazzling pattern (\textit{star design, diamond style...}) 
        & \textit{"pattern welcomed by superstar", "eyes-catching pattern", "pattern suitable for a dancing party"...} \\ \cmidrule(r){2-3}
        & modest pattern (\textit{stripes, checkered and plain...}) 
        & \textit{"simply decorated pattern", "pattern for the middle-aged", "pattern selected by humble people"...} \\ \midrule
        \multirow{4}*{\textbf{Material}} 
        & soft material (\textit{natural fibers, wool, leather...})  
        & \textit{"easily bent material", "gentle material", "material suitable for a child", "comfortable material"...} \\ \cmidrule(r){2-3}
        & gorgeous material (\textit{leather, wool, silk...}) 
        & \textit{"luxury material", "material indicating social status”, "material welcomed by rich people"...} \\ \cmidrule(r){2-3}
        & reliable material (\textit{metal, marble, plastic...}) 
        & \textit{"durable material", "material for restaurant furniture", ”material for outdoor furniture"...} \\
        \bottomrule
    \end{tabular}}
    \caption{Examples of some categorization concepts and corresponding subjective preferences.}
    \vspace{-0.5cm}
    \label{tab:preference}
\end{table*}

\section{SURE Dataset}
We build SURE (\textit{Multimodal Recommendation Dialog with \textbf{SU}bjective P\textbf{RE}ference} ) dataset to facilitate research on multimodal recommendation agents. In SURE dialogs, customers express their preferences subjectively. To effectively recommend items, the salespersons perform: \ding{182} Actively elicit customer preferences about item attributes; \ding{183} Disambiguate subjective preferences according to the multimodal context; \ding{184} Narrow the candidate set of items based on dialog history and scene; \ding{185} Recommend target item from the situated scene. 

To collect SURE dialogs, we design a two phrases pipeline (Fig.~\ref{fig:annotation}) following popular machine $\leftrightarrow$ human collaborative dialog collection approaches~\cite{simmc2, DBLP:conf/aaai/RastogiZSGK20, DBLP:journals/corr/abs-1801-04871}. In this section, we will introduce the SURE two phrases collection process in order and then analyze the distinguishing features of SURE.

\subsection{Multimodal Dialog Flow Simulation}
To generate dialog flows between salesperson and customer, we collect real-life shopping information first and then construct simulators based on this information. 

\subsubsection{Real-life Information Collection}
We invite human annotators to label scene background items and experienced sales experts to complete questionnaires about attribute categorization and multimodal recommendation strategy. 

\paragraph{Store Scene \& Background Item Annotation}
We develop SURE based on 1566 scene snapshots in SIMMC 2.0~\cite{simmc2}. These snapshots come from 140 fashion stores and 20 furniture stores generated by Unity 3D. To utilize spatial relations between commodity items and background items to facilitate recommendations, we invite Amazon Mechanical Turk (AMT) annotators with a higher than 95\% HIT approval rate to label bounding boxes of background items in fashion scenes (Appendix \ref{sssec:bg_item}). These background items include \textit{display table}, \textit{wardrobe}, \textit{floor rack} and \textit{display wall}. Each bounding box of background item covers all clothes in it and is labeled as \textit{"absolute position + name"} like \textit{"back leftmost closet"}. To ensure unambiguity, there are no repeated background item labels in the same snapshot. 

\paragraph{Attribute Categorization}
There are 290 and 110 different digital items in clothes and furniture domains. Nine attributes (e.g., \textit{type, color, pattern, material, price, brand, size and customer review}) are used for describing the items in metadata (the database of all digit items). 

Customers usually are not experts in those domains. They tend to express their requirements with what we call subjective preferences. On the one hand, a subjective preference can normally be mapped to a set of attribute values. For example, \textit{"color for happy activity"} corresponds to \textit{"red, brown, yellow, ..."}. On the other hand, there are lots of subjective preferences with the same meaning. For example, customers also say "color for the welcome ceremony" and "color of passion", which is similar to "color for happy activity" in attribute reference. The relationship between subjective preferences and attribute values is a many-to-many mapping. To bridge them, we collect a set of categorization concepts from the survey for domain experts(Appendix~\ref{sssec:survey}), each of which is a synonym of corresponding subjective preferences. Every subjective preference can be mapped to a categorization concept. Therefore, we transform the many-to-many mapping between subjective preferences and attribute values into two stages: \ding{182} Many-to-one mapping from subjective preferences to categorization concepts, \ding{183} One-to-many mapping from concepts to attribute values. 

The introduction of categorization concepts is necessary and convenient for the two phases TOD data building: In the dialog flow simulation, the customer simulator expresses requirements by categorization concepts as the slots. In the manual paraphrase, all categorization concepts are paraphrased to subjective preferences by human annotators. In this way, we can simulate dialog flow and guarantee language diversity at the same time.

Tab.~\ref{tab:preference} gives some examples of subjective preferences, categorization concepts, and attribute values. Fig.~\ref{fig:redial_tree} in Appendix show more details about them. 

\begin{table*}
    \centering
    \scalebox{0.6}{
    \begin{tabular}{ccl}
        \toprule
         {\textbf{\textsc{Salesperson Act}}} & {\textbf{\textsc{Customer Act}}} & \textbf{\textsc{Examples}}   \\
        \midrule
         \multirow{2}*{Ask Preference}  & \multirow{2}*{Answer Preference} 
         & A:\textit{"Could you please tell me your preference on clothes pattern?"} \\ 
         & & $\rightarrow$ U:\textit{"I would like to choose clothes with \textbf{pattern that is suitable for my dancing party}."} \\
         \midrule
         \multirow{2}*{Exclude Preference} & \multirow{2}*{Negate Preference}  
         & A:\textit{"Is there any kind of furniture material you don't like? I can avoid recommending it to you."} \\
         & & $\rightarrow$ U:\textit{"I will not choose \textbf{materials popular among rich people}. I am used to a simple life." }  \\
         \midrule
         \multirow{2}*{Prompt Preference} & \multirow{2}*{Respond Prompt}
         & A:\textit{"In this case, what is your opinion on the \textbf{color that makes people feel cool}?"} \\ 
         & & $\rightarrow$ U:\textit{"Cool color sounds Great! It is suitable for my summer vacation. Just follow your suggestion."}\\
         \midrule
         \multirow{2}*{Guess Attribute Value} & \multirow{2}*{Respond Attribute Value} 
         & A:\textit{“Can you tell me how do you think about \underline{leather material}?”} \\                                     
         &  & $\rightarrow$ U:\textit{"I like this kind of material. Just recommend leather furniture to me later."} \\   
         \midrule
         \multirow{2}*{Revise Attribute Value}  & \multirow{2}*{Respond Attribute Value} 
         & A:\textit{"How about brand \underline{Art News Today}?"} $\rightarrow$ U:\textit{"I don't like \textbf{brand with a long history}."} \\
         & & (next round) A:\textit{"I see .. would you like trademark \underline{Coats \& More}?"} $\rightarrow$ U:\textit{"Oh! You know my taste."} \\
         \midrule
         \multirow{2}*{Display Candidate Values} & \multirow{2}*{Choose Attribute Value}  
         & A:\textit{“I have jackets with \underline{blue color, green color and white color}. Which color do you prefer?”} \\  
         & & $\rightarrow$ U:\textit{"Let me have a think. Oh, as for me, I would like \underline{white color} for jacket}."  \\
         \midrule
         \multirow{2}*{Refer Region} & \multirow{2}*{Judge Region}  
         & A:\textit{"Have a look at the \textbf{back left wardrobe near the mirror}. Is there anything you like in this region?"}  \\
         & & $\rightarrow$ U:\textit{"Sorry, there seem to be no suitable clothes in this closet."} \\
         \midrule
         \multirow{2}*{Recommend Item} & \multirow{2}*{Respond Recommendation}  
         & A:\textit{"What is your idea of the black and white t-shirt with stripes hanging on the \textbf{rightmost floor rack}?"} \\
         & & $\rightarrow$ U:\textit{"Yeah, it looks very beautiful! Please help me add this t-shirt to my cart."} \\
        \bottomrule
    \end{tabular}}
    \caption{Sixteen kinds of salesperson acts and customer acts in SURE dialogs. Corresponding dialog examples for each pair of acts are displayed in the last column. All subjective preferences and referred regions are highlighted in bold while concrete attribute values are underlined.}
    \vspace{-0.5cm}
    \label{tab:strategy}
\end{table*}

\paragraph{Salesperson Act.}
To collect common salesperson acts, we invite 238 sales experts with more than three years of work experience to complete questionnaires (Appendix~\ref{sssec:survey}). The maximum entry number is limited to 1 to guarantee result diversity. We summarize 8 different dialog acts of salespersons (Tab.~\ref{tab:strategy}) and introduce them in detail.

\begin{itemize}
    \item[$\bullet$] \textbf{\textit{Ask Preference}} refers to asking customer's preference about one attribute type. Overlapping values referred to by all subjective preferences responded from the customer are new candidates. 
    \item[$\bullet$] \textbf{\textit{Exclude Preference}} is asking the customer what he dislike. The responded subjective preference from customer is utilized to exclude unfavored attribute values from candidates. 
    \item[$\bullet$] \textbf{\textit{Prompt Preference}} is actively providing subjective "preference" for the customer to confirm. 
    \item[$\bullet$] \textbf{\textit{Guess Attribute Value}} is predicting one concrete value from candidate attribute values based on multimodal context.
    \item[$\bullet$] \textbf{\textit{Revise Attribute Value}} is revising the previous prediction of concrete attribute value following the customer's feedback.
    \item[$\bullet$] \textbf{\textit{Display Candidate Values}} refers to listing all candidate attribute values based on multimodal context for the customer to choose from. 
    \item[$\bullet$] \textbf{\textit{Refer Region}} is an act that salesperson points out one region in the store like \textit{"front floor rack"} to ask the customer whether the region contains the item he wants. 
    \item[$\bullet$] \textbf{\textit{Recommend Item}} is trying to recommend items from the candidate item set based on multimodal context. 
\end{itemize}

Different combinations of the above acts in one dialog can form diverse recommendation strategies. For example, a salesperson can continually ask customer preferences to narrow the candidate set of attribute values, and then display all possible values for the customer to choose when several attribute values are in the candidate set. Using reasonable strategies for different situations can effectively improve recommendation efficiency and accuracy.

\subsubsection{Multimodal Dialog Simulator}
The multimodal dialog simulator takes store scenes along with the meta information to create salesperson-customer dialog flows following~\cite{simmc2}.

\paragraph{Multimodal Dialog Flow Generation.}
The dialog flow simulator is composed of the \textit{goal generator}, the \textit{customer simulator} and the \textit{salesperson simulator}. The goal generator randomly selects an item from the given store scene and takes it as a target item that is invisible to others. The salesperson simulator is aware of the scene snapshot (commodity and background items' position) and metadata of all items, which actively elicits customer preferences and recommends items following a probability distribution. Each act (e.g.,~\textit{Prompt\_Reference, Refer\_Region}) is companied by some slots (e.g.,~attribute, region). The customer simulator is assigned attribute values of the target item and responds to salespersons by customer acts with slot values limited to categorization concepts or simple yes/no (except \textit{Display\_Candidate\_Values}), which simulates the user requirements. Take one-round interaction as an example. After the salesperson simulator generates \textit{"Ask\_Preference}:\{Color\}\textit{"}, the customer simulator chooses its act \textit{"Answer\_Preference"} and a categorization concept \textit{"warm color"} based on assigned \textit{"red"} color as the preference slot. The simulation repeats until the salesperson simulator successfully recommends the target item.

\begin{figure*}[ht]
    \centering
    \includegraphics[width=1\linewidth]{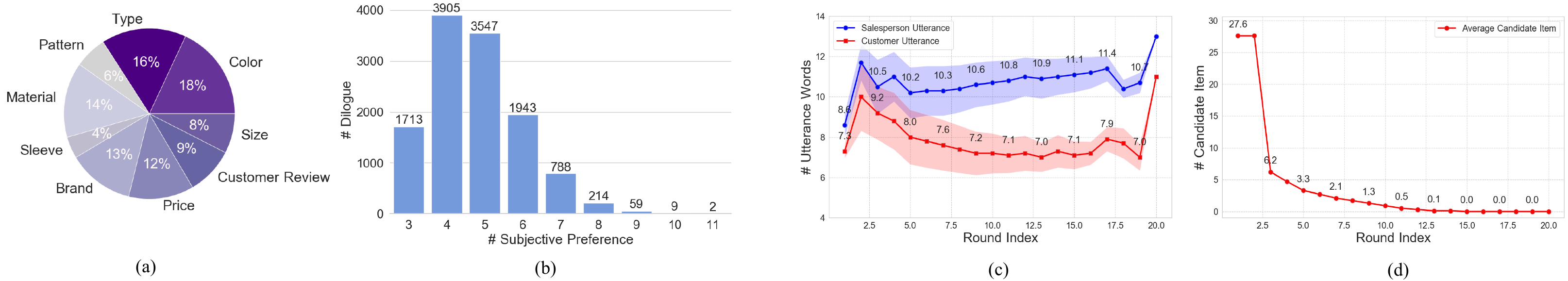}
    \caption{(a) Percentage of subjective preferences in attribute types, (b) distribution of subjective preferences among dialogs, (c) number of utterance words with dialog rounds, (d) number of candidate items with dialog rounds.}
    \vspace{-0.3cm}
    \label{fig:statistic}
\end{figure*}

\begin{figure*}[ht]
    \centering
    \includegraphics[width=1\linewidth]{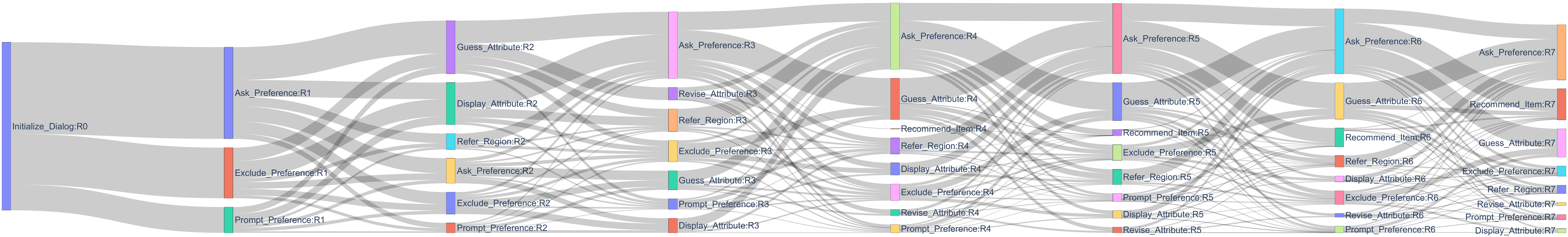}
    \caption{Salesperson act transitions in the first eight rounds. Different combinations of salesperson acts form diverse strategies to recommend the target item from complex store scene based on subjective preferences.}
    \vspace{-0.3cm}
    \label{fig:sankey}
\end{figure*}

\subsection{Manual Paraphrase}
Based on dialog flows obtained from simulator interaction, we design a manual paraphrase process to rewrite subjective preferences and then paraphrase dialog flows.

\paragraph{Subjective Preference.}
In simulated dialog flows, subjective preferences are expressed by categorization concepts for attribute values. Every categorization concept can be paraphrased to many different subjective preferences. Take \textit{"warm color"} as an example. It can be rewritten to \textit{"color of passion"} by customers' subjective feeling, to \textit{"color welcomed by outgoing people"} by suitable persons, to \textit{"color for the happy ceremony"} by applicable scenarios. Human annotators are required to paraphrase categorization concepts to subjective preferences in any of the cases to increase language diversity.

\paragraph{Dialog Generation.}
To make dialogs closer to language distribution of real shopping dialogs, we invite AMT annotators with a higher than 90\% HIT approval rate to paraphrase dialog flow following these instructions: \ding{182} Write salesperson $\leftrightarrow$ customer utterances based on dialog flows. All subjective preferences and concrete attribute values are reserved; \ding{183} Add the visual descriptions and spatial relations of scene items; \ding{184} Rewrite repeated nouns or phrases to co-reference; \ding{185} Add polite expressions and modal particles into utterances. The detailed instructions with scene snapshot and dialog flow can be checked in Appendix~\ref{sssec:paraphrase}. An example of SURE dialog with annotations is shown in Appendix Fig.~\ref{fig:dialog_case}.

\subsection{SURE Dataset Analysis}
To the end, we build the SURE dataset, Tab.~\ref{tab:statistic} gives some statistics of the data. We highlight the information on subjective preference and dialog policy in the following subsections.

\begin{table}[h]
    \centering
    \scalebox{0.6}{
    \begin{tabular}{lc}
        \toprule
         Total \# dialogs     & 12180 \\
         Total \# utterances   & 223492  \\
         Total \# scene snapshots  & 1566  \\
         Total \# subjective preferences  & 3043 \\
         Avg \# words per customer turns & 7.75  \\
         Avg \# words per salesperson turns  & 10.49  \\
         Avg \# utterances per dialog  & 18.35  \\
         Avg \# objects per scene in dialog  & 27.6 \\
         Avg \# subjective preferences per dialog  &  4.48  \\ 
         Avg \# salesperson acts per dialog &  8.17  \\
        \bottomrule
    \end{tabular}}
    \caption{SURE Dataset Statistics.}
    \vspace{-0.5cm}
    \label{tab:statistic}
\end{table}

\paragraph{Subjective Preference.}
As shown in Tab.~\ref{tab:statistic}, there are 3$K$ different subjective preferences in SURE. The percentage of subjective preferences in different attribute types is shown in Fig.~\ref{fig:statistic}~(a), from which we can observe the richness and diversity of subjective preferences. On average, each dialog contains 4.48 subjective preferences. The distribution of subjective preferences among dialogs is displayed in Fig.~\ref{fig:statistic}~(b). It is clear that subjective preferences are widely distributed in SURE, which brings a new challenge for conversational agents. 

\begin{figure*}[t]
    \centering
    \includegraphics[width=0.65\linewidth]{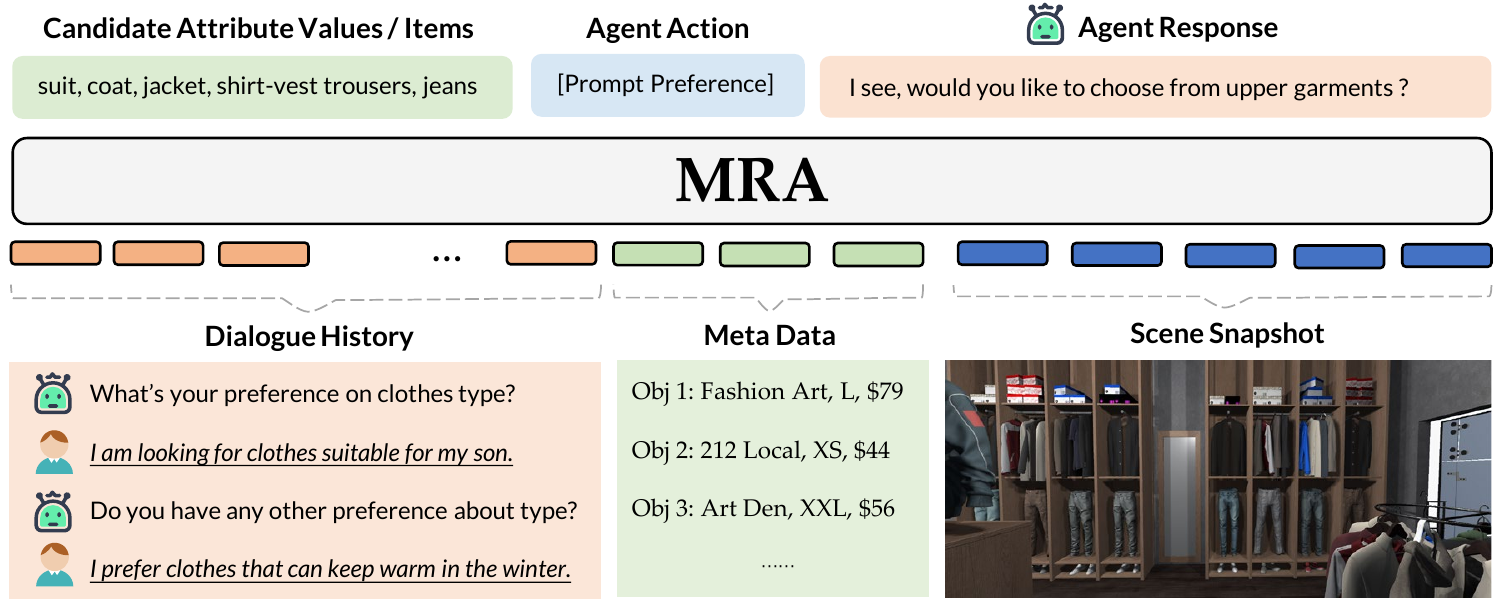}
    \caption{The MRA model for \textbf{Subjective Preference Disambiguation}, \textbf{Referred Region Understanding} and \textbf{Multimodal Recommendation} tasks. The flattened dialog history and metadata are concatenated as textual input while the scene snapshot is split into patches. Note that the green area in the output will predict candidate attribute values if the latest round relates to eliciting preference. When the latest round refers to specific region, the green area predicts object IDs in the region.}
    \vspace{-0.5cm}
    \label{fig:baseline}
\end{figure*}

\paragraph{Dialog Policy.}
As shown in Tab.~\ref{tab:statistic}, the SURE dataset collects 12$K$ shopping recommendation dialogs. The utterance length with dialog turn is displayed in Fig.~\ref{fig:statistic}~(c). On average, each dialog contains 8.17 salesperson acts to recommend the target item from 27.6 candidate scene items. We visualize the salesperson act transactions for the first eight rounds in Fig.~\ref{fig:sankey}. It can be observed that different act combinations form diverse recommendation strategies in SURE. From the stream width, we can find that salespersons have a higher probability of asking for preference than excluding and prompting preferences. Besides, salespersons don't directly guess or display concrete attribute values at the very beginning round. They are prone to conduct these acts after they collect at least one customer preference. As the dialogs go on, salespersons also try to reduce candidate items by referring to specific region. We display the number of candidate items over rounds in Fig.~\ref{fig:statistic}~(d). It can be seen that salespersons begin to recommend items when the candidate range is small enough. The recommendation strategies in SURE are close to real-life shopping.

\section{Task Formulation}
We propose three tasks on SURE dataset to evaluate multimodal recommendation agents. The tasks of Subjective Preference Disambiguation and Referred Region Understanding evaluate the side of multimodal understanding, while Multimodal Recommendation evaluates the side of policy learning.  

\paragraph{Subjective Preference Disambiguation (SPD).}
After the customer expresses subjective preferences, the agent needs to determine candidate attribute values based on preferences in dialog history and scene items in the store. We denote this process as subjective preference disambiguation, which establishes the connection between subjective preferences and concrete attribute values. This task is important because the correct recommendation is closely dependent on clear attribute requirements. It requires the agent to abstract subjective preferences to categorization concepts and then filter grounded attribute values by these concepts. From cognition research, this task involves visual perception, language conceptualization and attribute categorization.

The input of this task includes dialog history, current customer utterance, and scene snapshot. With this information, the agent predicts all possible attribute values (e.g., U:\textit{"I prefer the color of happiness."} $\rightarrow$ \textit{"yellow, brown, red"}). Attribute values that meet customer requirements but do not exist in the scene snapshot should be excluded. The main evaluation metric can be F1, precision, and recall performance. Note that the evaluation is only implemented on rounds for eliciting preferences.

\paragraph{Referred Region Understanding (RRU).}
Referring to region is an essential act for narrowing the candidate item range. This task aims to update the candidate item set after the customer responds to referred region. To complete this task, the agent needs to locate the regional referring expression (e.g., \textit{"far back middle floor rack"}) and then filter the previous candidate set by items in the region. It requires the agent to correctly understand the referred region, visual attributes and spatial relations in the scene. 

The input of this task is dialog history with the latest round containing referred region and scene snapshot. Based on this information, the agent needs to predict all object IDs in the region (e.g., A:\textit{"Come with me to look at the shelf on the right. Are there any clothes that you like?"} U:\textit{Sorry, there is no garment that I am looking for in this region.} $\rightarrow$ \textit{"12, 13, 16, 22, 31"}). Objects in the same scene but not in the referred region should be excluded. The agent performance can be measured by F1 score, precision, and recall metrics on object ID prediction. Note that the evaluation is only implemented on rounds for referring region. 

\begin{table*}[t]
    \centering
    \scalebox{0.7}{
    \begin{tabular}{l|c|c|ccc}
        \toprule
        \multirow{2}*{\textbf{\textsc{Models}}} & \textbf{Task 1 SPD} & \textbf{Task 2 RRU} & \multicolumn{3}{c}{\textbf{Task 3 MR}}  \\
        \cmidrule(r){2-6}
         ~ & Disam. F1$\uparrow$ & Refer F1$\uparrow$ & Act F1$\uparrow$  & BLUE-4$\uparrow$ & Recom. F1$\uparrow$ \\
        %\cmidrule(r){1-2} \cmidrule(r){3-3} \cmidrule(r){4-4} \cmidrule(r){4-7}
        \midrule
        MRA  & \textbf{36.78 / 37.49} & \textbf{13.77 / 14.77} & \textbf{10.80 / 10.90} & \textbf{18.28 / 18.45} & \textbf{23.49 / 23.14}  \\
        - Meta Data & 32.34 / 31.98   & 9.57 / 10.15   & 10.01 / 10.21   & 16.42 / 16.83   & 19.78 / 19.23   \\
        - Scene Snapshot & 31.44 / 31.83  & 7.36 / 7.09   &  9.45 / 9.77   & 17.72 / 17.63  & 21.54 / 21.26   \\
        \bottomrule
    \end{tabular}}
    \caption{The performance of baseline model MRA on three SURE benchmarks. MRA's results on dev-test and test-std are displayed by \textbf{"(dev-test / test-std)"} format. We respectively ablate metadata and scene snapshot to observe MRA's ability to utilize multimodal context to recommend item via subjective preferences.}
    \vspace{-0.5cm}
    \label{tab:result}
\end{table*}

\paragraph{Multimodal Recommendation (MR).}
When customers seek recommendations in the store, they hope the salesperson can recommend items accurately and efficiently. Therefore, recommendation strategy, recommendation success rate, and language quality all influence customer shopping experience. We define \textbf{Act Prediction} sub-task and \textbf{Response Generation} sub-task at turn level and define \textbf{Item Recommendation} sub-task at dialog level to evaluate the agent's multimodal recommendation performance comprehensively.

The input of these three sub-tasks is dialog history, current customer utterance, and scene snapshot. The \textbf{Act Prediction} sub-task requires the agent to predict the next salesperson act (e.g., U:\textit{"The price \$299 is too expensive for me to afford" $\rightarrow$ "Revise\_Attribute"}). The F1 score, precision, and recall can be calculated for cumulative act predictions to measure performance. The \textbf{Response Generation} sub-task requires the agent to generate the next salesperson utterance (e.g., U:\textit{"I'd like to buy a sofa made by materials obtained from nature" $\rightarrow$ "You can consider the leather material, which is natural and smooth."}). The generated utterance can be evaluated by BLEU-4~\cite{bleu} or ROUGE~\cite{rouge}. The \textbf{Item Recommendation} sub-task requires the agent to predict the target item ID (e.g., \textit{"<@1132>"}) in the last round utterance, which can be extracted by regex and evaluated by F1 score, precision, and recall. 

\section{Modeling \& Empirical Analysis}
\paragraph{Dataset Split.}
SURE is randomly divided into 4 parts: train (65\%), dev (5\%), dev-test (15\%), and test-std (15\%). We leave test-std as a held-out hidden set for performing a fair comparison of models in future potential competition.

\paragraph{Baseline.}
We proposed MRA(Multimodal Recommendation Agent) model as the baseline model for subjective preference disambiguation, referred region understanding, and multimodal recommendation tasks. The backbone of the MRA model is encoder-decoder based single-stream Visual-Language Pre-training Model, which are stacks of Transformer~\cite{transformer} layers. The scene image is split into $P$ patches. And each patch is projected to the visual embedding of the model's hidden size. The flattened dialog history and non-visual metadata are converted to sub-word sequences by Byte-Pair Encoding (BPE) and then embedded into textual embedding. All visual embedding and textual embedding are concatenated as model input. The MRA model completes three benchmarks at the same time by generating \textit{next salesperson act, candidate attribute values\,/\,referred region items} and \textit{agent response} auto-regressively as Fig.~\ref{fig:baseline} shows. Note that MRA will predict candidate attribute value if the latest round relates to eliciting preference. When the latest round relates to referring region, MRA predicts object IDs in the region.

\paragraph{Implementation Details.}
MRA model is based on Transformer~\cite{transformer} structure with 12 layers, where every Transformer block has 768 hidden units and 12 attention heads. Textual and visual embedding are projected to features the same size as the hidden units. We initialize MRA parameters from pretrained OFA-base~\cite{ofa} model. During training, MRA model is trained for 20 epochs with 18 batch sizes to optimize language modeling loss. At the end of every epoch, the model is evaluated on dev split to save the best model parameters. The hyperparameters are determined by area search. Adam~\cite{adam} is adopted as the optimizer with a 4e-4 learning rate while the dropout rate is set to 0.2 to prevent over-fitting. The whole training costs around 36 Tesla-V100 GPU hours. Note that the BLEU-4 score is calculated by NLTK~\cite{nltk} package in the evaluation.

\paragraph{Analysis and Future Work}
As Tab.~\ref{tab:result} shows, model MRA, powered by an advanced multimodal backbone, fails to perform well on three SURE benchmarks. From the case study (Fig.~\ref{fig:case_study}), we can observe that it is difficult for MRA to accurately understand subjective preferences (Task 1) and referred region (Task 2), which further hinders the model from adopting suitable acts to make correct recommendations (Task 3). Ablation of metadata and scene snapshot greatly weakens MRA on the first two tasks. It indicates that effective utilization of metadata and scene snapshot plays an essential role in model performance. Future work can be done by designing modules or proposing multimodal pretraining tasks to facilitate model's understanding of subjective preferences, perception of referred region, and ability to take suitable acts.

\section{Conclusion}
We introduce \textit{Multimodal Recommendation Dialog with \textbf{SU}bjective P\textbf{RE}ference} (SURE) dataset with 12$K$ salesperson $\leftrightarrow$ customer dialogs and 3$K$ subjective preferences to study how to recommend item from complex scene based on subjective preferences. Our proposed three benchmarks and strong baseline model MRA address the new challenges and directions in the multimodal recommendation dialog.

\section*{Limitations}
The annotation of attribute categorization and subjective preferences may vary from person to person, influencing preference disambiguation results in the real world. We have tried to reduce bias by choosing categorization concepts and subjective preferences agreed upon by more than three annotators. Besides, owing to time and funds constraints, we only manually paraphrase dialog flow in English. For this reason, the agent built on SURE can just communicate in English. To overcome this limitation, we plan to annotate SURE in multi-language in the next stage.

\section*{Ethics Statement}
Our work strictly complies ACL Code of Ethics. We respect CC-BY-NC-SA-4.0 license required by SIMMC 2.0 and only use its scene snapshots for academic research. We will also release our dataset to the community with the same license. As for human annotation, we anonymously recruit human annotators on the Amazon Mechanical Turk (AMT) platform to protect their personal privacy. In the task instructions, we have informed participators that any annotations related to personal attacks, racial or sexism discrimination will lead to HIT rejection. Besides, we also demonstrate that their annotations will be used for academic purposes. The payment of our released tasks is competitive on the AMT platform compared with similar tasks (Appendix~\ref{sec:appendix}). After completing human annotations, we manually check the collected information to exclude any potential offensive information. Our annotation process and data content got approval from an ethics review board by an anonymous IT company. We can guarantee the trustworthiness of our technologies, limitations, and ethics statement.

\section*{Acknowledgements}
We would like to sincerely thank anonymous reviewers for their suggestions and comments. The work was partially supported by the National Natural Science Foundation of China (NSFC62076032). We also want to express our gratitude for precious advises given by Guanqi Zhan.

\bibliography{anthology,custom}
\bibliographystyle{acl_natbib}

\appendix

\section{Appendix}
\label{sec:appendix}

\subsection{Human Annotation}
\subsubsection{Background Item Annotation.} \label{sssec:bg_item}
We release "Background Item Annotation" task on Amazon Mechanical Turk (AMT) platform to hire workers to draw bounding box around background item and annotate corresponding label. To guarantee the quality, we require workers have greater than 95\% HIT approval rate. For payment, we pay \$0.25 for every store scene snapshots that contains about 5 potential background items, which is competitive compared with similar tasks. The detailed task instruction and scene snapshot are displayed in the following. \\

\centerline{\textit{\textbf{Background Item Annotation}}}
\textit{This HIT is a part of scientific research, whose results may be presented at scientific meetings or published in scientific journals. In this task, you are invited to draw all bounding boxes of background items, like floor rack, display table and shelf, in the given scene snapshot and annotate the corresponding label. The annotated bounding box has to cover all clothes bounding boxes contained in it and the label should be in the "absolute position + name" format such as "back leftmost closet". The hit will be rejected if the annotated bounding box does not cover all clothes bounding boxes contained in it or the label has racism, sexism and privacy information. If you are fully aware of and agree with above information, you can begin to work on the following scene snapshot.}

\begin{figure}[h]
    \centering
    \includegraphics[width=1\linewidth]{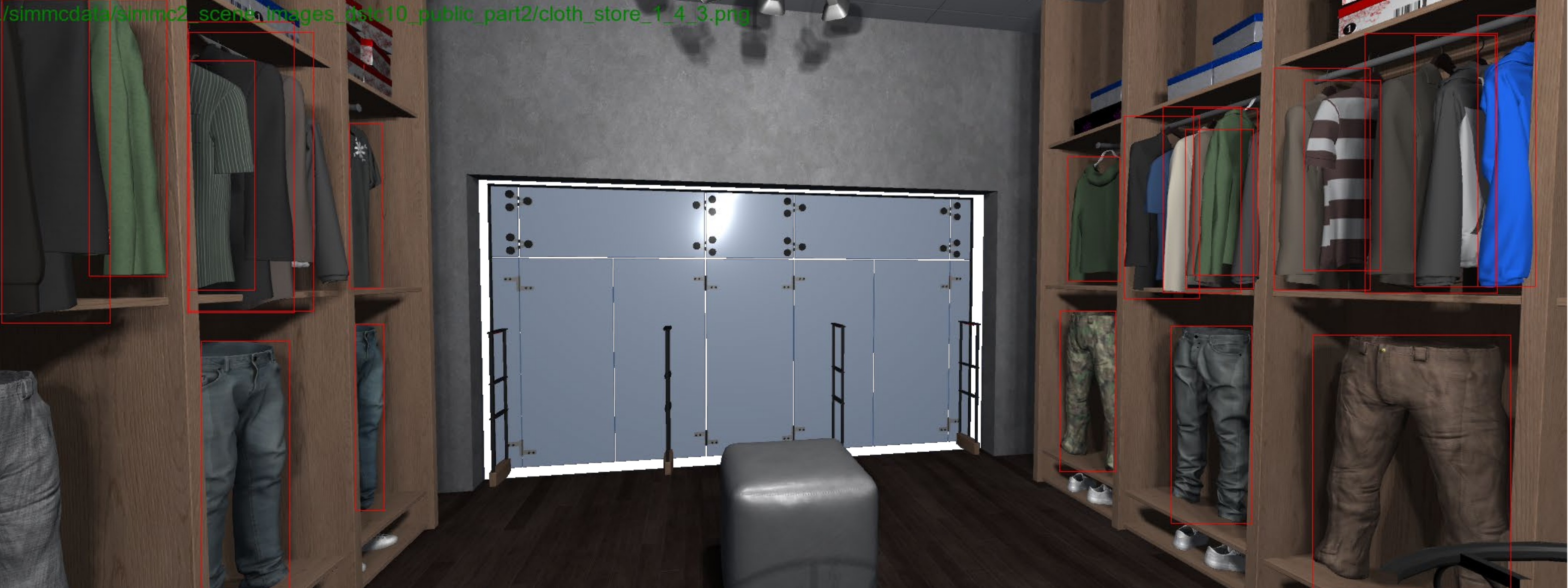}
    \caption{Scene snapshot of background item annotation. Existing red bounding boxes are for clothes items.}
    \label{fig:bg_item}
    \vspace{-0.3cm}
\end{figure}

\subsubsection{Questionnaire for Sales Expert} \label{sssec:survey}
We release "Clothes Recommendation Survey" task and "Furniture Recommendation Survey" task on Amazon Mechanical Turk (AMT) platform to invite fashion sales experts and furniture sales experts to complete questionnaire. To guarantee the quality, we require answers have greater than 90\% HIT approval rate. For payment, we pay for \$2.0 for every carefully completed questionnaires, which is high compared with other survey tasks in the same period. We display the instruction and questionnaire for "Clothes Recommendation Survey" task in the following. \\

\centerline{\textit{\textbf{Clothes Recommendation Survey}}}
\textit{Our project is aimed at studying clothes recommendation in the store scene. To collect real-life data, we sincerely invite experienced sales expert to complete the following questionnaire with 12 questions. *\,We guarantee that all questionnaires will be conducted anonymously, and the survey results will only be used for academic research rather than commercial purposes. The hit will be rejected if the comments contain any racism, sexism and privacy information. If you are fully aware of and agree with above information, you are welcome to accept the survey. \\ \\
\textbf{1.\;How long work experience do you have in clothing sales?} (Single Choice Question) \\
A.\;I don't have work experience on clothing sales. \\
B.\;Less than 1 year. \\
C.\;About 1 - 3 years. \\
D.\;About 3 - 5 years. \\
E.\;More than 5 years. \\ \\
\textbf{2.\;What order of priority do you usually follow when eliciting customer preferences on attributes?} (Sorting Question) \\
$\bullet$\;Type (e.g., jacket, t-shirt, dress) \\
$\bullet$\;Color (e.g., red, blue, white) \\
$\bullet$\;Pattern (e.g., plain, stripe, spot) \\
$\bullet$\;Size (e.g., XS, M, XXL) \\
$\bullet$\;Sleeve Length (e.g., short, full, half) \\
$\bullet$\;Brand \\
$\bullet$\;Price \\ \\
\textbf{3.\;Do you feel how often customer express their requirements on attribute value by subjective preference such as "color of passion", "lively pattern" and "formal fashion type"?} (Single Choice Question) \\
A.\;More than 60\% of time. \\
B.\;About 40\% - 60\% of time. \\
C.\;About 20\% - 40\% of time. \\
D.\;Less 20\% of time. \\ \\
\textbf{4.\;Please categorize the given attribute values by your domain knowledge and subjective feeling. For example, “red, yellow, brown...” can be categorized to "warm" color while "floral pattern, leopard print..." can be categorized to "lively pattern". You can just select several values from candidates to define categorization concept. Then, write down some subjective preference your customer expressed if you have met such case.} (Leave Comments) \\
Given Attribute Values: red, yellow, brown, orange, pink, blue, black, grey, dark olive, light red. (These provided attribute values vary from questionnaire to questionnaire.) \\ \\
\textbf{5.\;Do you think knowing about what kinds of concrete attribute values are most important for successful recommendation?} (Multiple Choice) \\
A.\;Type (e.g., jacket, t-shirt, dress) \\
B.\;Color (e.g., red, blue, white) \\
C.\;Pattern (e.g., plain, stripe, spot) \\
D.\;Size (e.g., XS, M, XXL) \\
E.\;Sleeve Length (e.g., short, full, half) \\
F.\;Brand \\
G.\;Price \\ \\
\textbf{6.\;When customer express subjective preferences, what strategy will you utilize to disambiguate their requirements? For example, "color of passion"$\rightarrow$"red, orange, yellow...".} (Multiple Choice) \\
A.\;Continually ask customers more preferences on this attribute type. \\
B.\;Invite customers to describe some information about their unfavored attribute values. \\
C.\;Actively prompt some subjective preferences on attribute type to inspire customers. \\
D.\;Directly display all attribute values in the candidate set for customers to choose. \\
E.\;Guess one concrete attribute value based on customer expressed preferences and revise prediction following feedback. \\
F.\;You are welcome to tell us your personal strategy! (Leave Comments) \\ \\
\textbf{7.\;At the time you elicit customer preference on one particular clothes attribute, how many attribute values remain in the candidate attribute set when you list them for customer to choose?} (Single Choice Question) \\
A.\;More than 8 candidate attribute values. \\
B.\;About 5 - 8 candidate attribute values. \\
C.\;About 3 - 5 candidate attribute values. \\
D.\;About 1 - 3 candidate attribute values. \\
E.\;Only 1 candidate attribute value. \\ \\
\textbf{8.\;Collecting information will promote recommendation accuracy but may consume customer patience. In this case, how do you balance collecting customer preference and trying recommending items?} (Single Choice Question) \\
A.\;I will learn about customer preference about all clothes attributes in detail first, and recommend item when having full confidence. \\
B.\;I will learn about customer preference roughly first, and recommend item when candidate range is small enough. \\
C.\;I will learn about detailed customer preference on 1 - 2 clothes attributes, and then try to recommend item until customer accept. \\
D.\;I will learn about detailed customer preference on 3 - 4 clothes attributes, and then try to recommend item until customer accept. \\ \\
\textbf{9.\;When will you guide customers to a particular region and invite them to see if there are suitable clothes? For example, "Please follow me to have a look at the \underline{left display wall}. Are there anything you like?".} (Single Choice Question) \\ 
A.\;After learning about detailed customer preferences on 1 - 2 clothes attributes, I will select one rack in the store and show it to the customer. \\
B.\;After learning about detailed customer preferences on 3 - 4 clothes attributes, I will select one rack in the store and show it to the customer. \\
C.\;Only when the range of candidate items is small enough, I will select one rack in the store and show it to the customer. \\
D.\;I don't do like what the question says. I always directly point out clothes which customer may be interested in. \\ \\
\textbf{10.\;How many clothes items remain in the candidate item set when you recommend concrete item for customer?} (Single Choice Question) \\
A.\;More than 8 candidate items. \\
B.\;About 5 - 8 candidate items. \\
C.\;About 3 - 5 candidate items. \\
D.\;About 1 - 3 candidate items. \\
E.\;Only 1 candidate item. \\ \\
\textbf{11.\;How do you unambiguously refer the item in the store when you recommend item to customer?} (Single Choice Question) \\
A.\;Refer the item by its color. \\
B.\;Refer the item by its pattern, \\
C.\;Refer the item by its color and pattern. \\
D.\;Refer the item by spatial relation. \\
E.\;Combine all above methods to refer item although its description is complex. \\ \\
\textbf{12.\;What do you think is the most important when shopping recommendation dialogs?} (Single Choice Question) \\
A.\;Recommend customer desired item through as few conversations as possible. (i.g. recommendation efficiency) \\
B.\;Keep recommendation strategies be consistent with those in real-life shopping conversations. \\
C.\;Balance the recommendation efficiency and customer shopping experience. \\
E:\;Guarantee dialog language polite and warm. \\
F:\;You are welcome to write your advice! (Leave Comments) 
}\\

\begin{figure}[htp]
    \centering
    \includegraphics[width=0.7\linewidth]{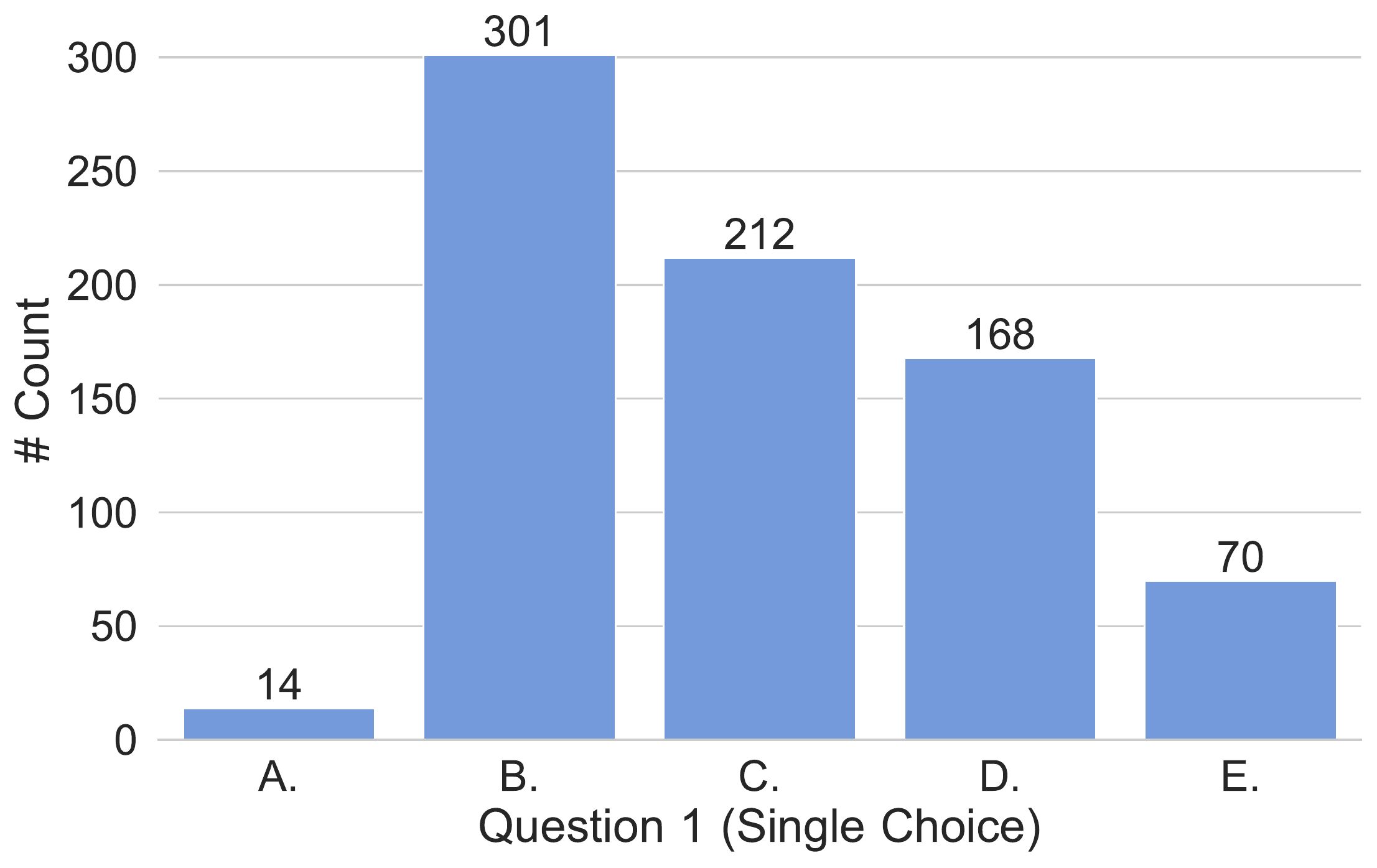}
    \caption{The bar chart of Question 1 results. \textbf{Choice D} and \textbf{Choice E} are selected by 238 sales expert with more than three years work experience. Only these 238 experienced sales expert are invited to complete the following Question 2 - Question 12.}
    \label{fig:question_1}
    \vspace{-0.4cm}
\end{figure}

\begin{figure}[htp]
    \centering
    \includegraphics[width=0.5\linewidth]{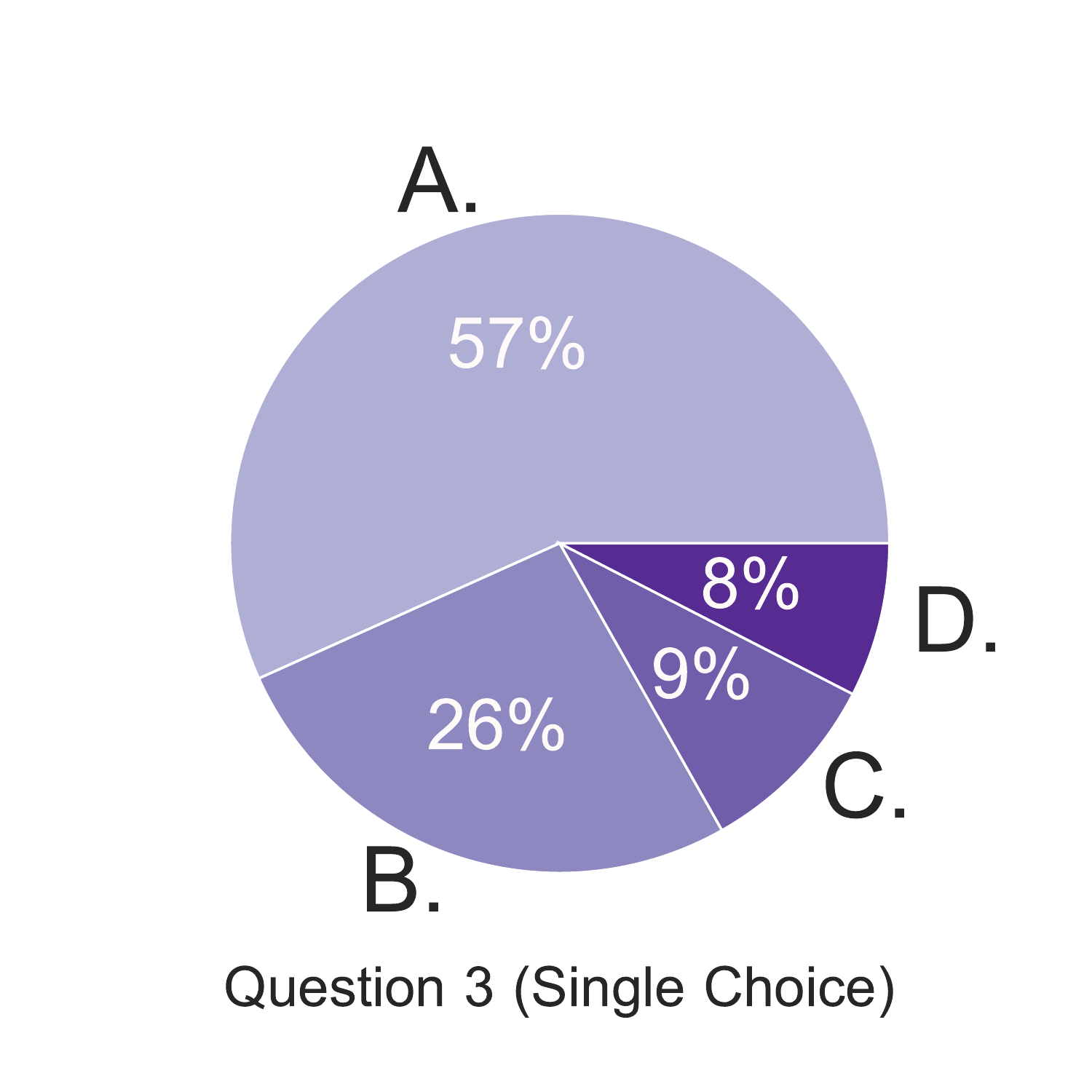}
    \caption{The pie chart of Question 3 results. From \textbf{Choice A} selection (57\%), we can find that over half of sales expert often meet customers who express their requirements on attribute value by subjective preferences.}
    \label{fig:question_3}
    \vspace{-0.4cm}
\end{figure}

\begin{figure}[htp]
    \centering
    \includegraphics[width=0.7\linewidth]{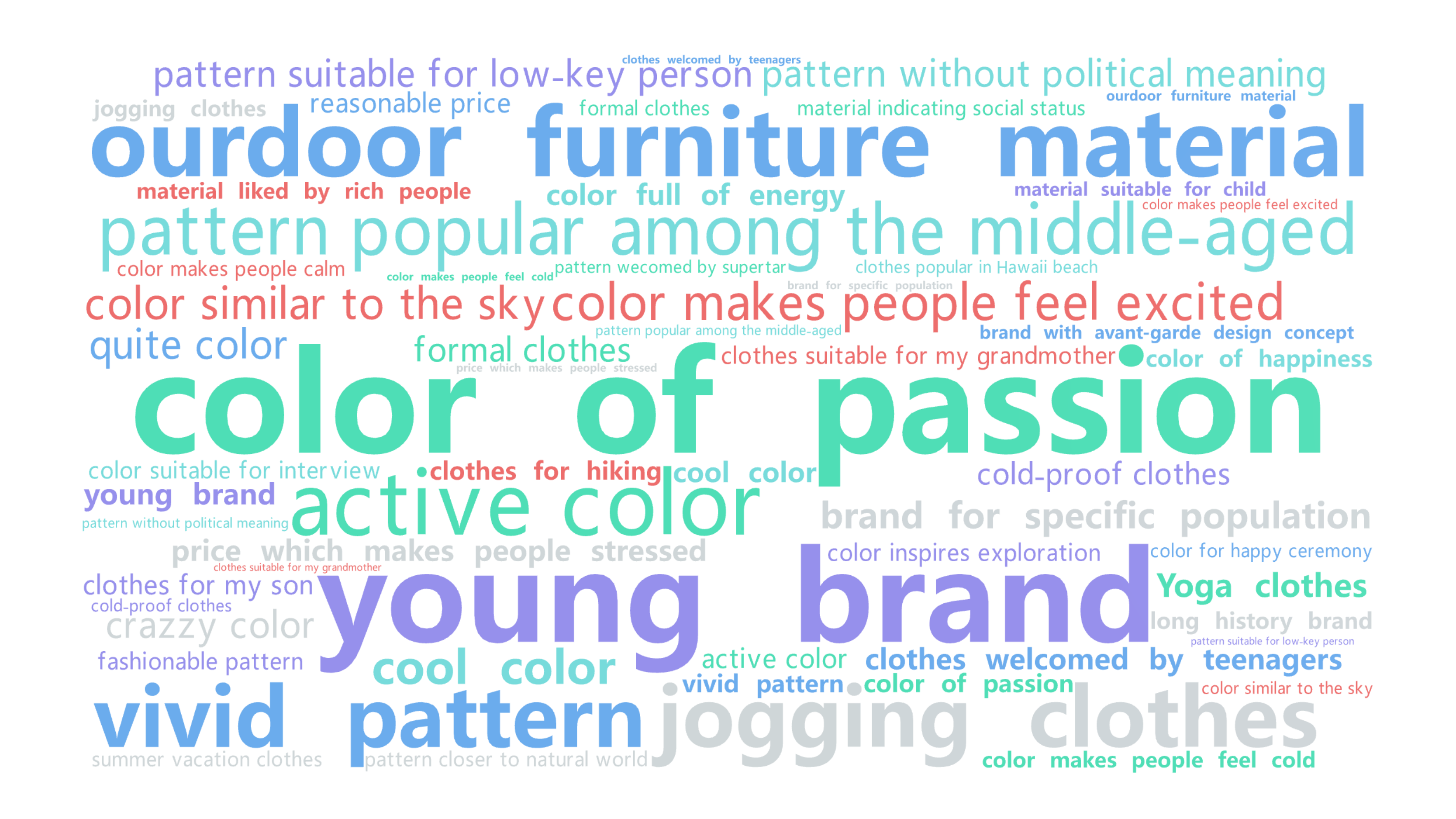}
    \caption{The word cloud of Question 4 comments. We can find that customers usually express subjective preferences by visual perception and commonsense.}
    \label{fig:question_4}
    \vspace{-0.4cm}
\end{figure}

\begin{figure}[htp]
    \centering
    \includegraphics[width=0.7\linewidth]{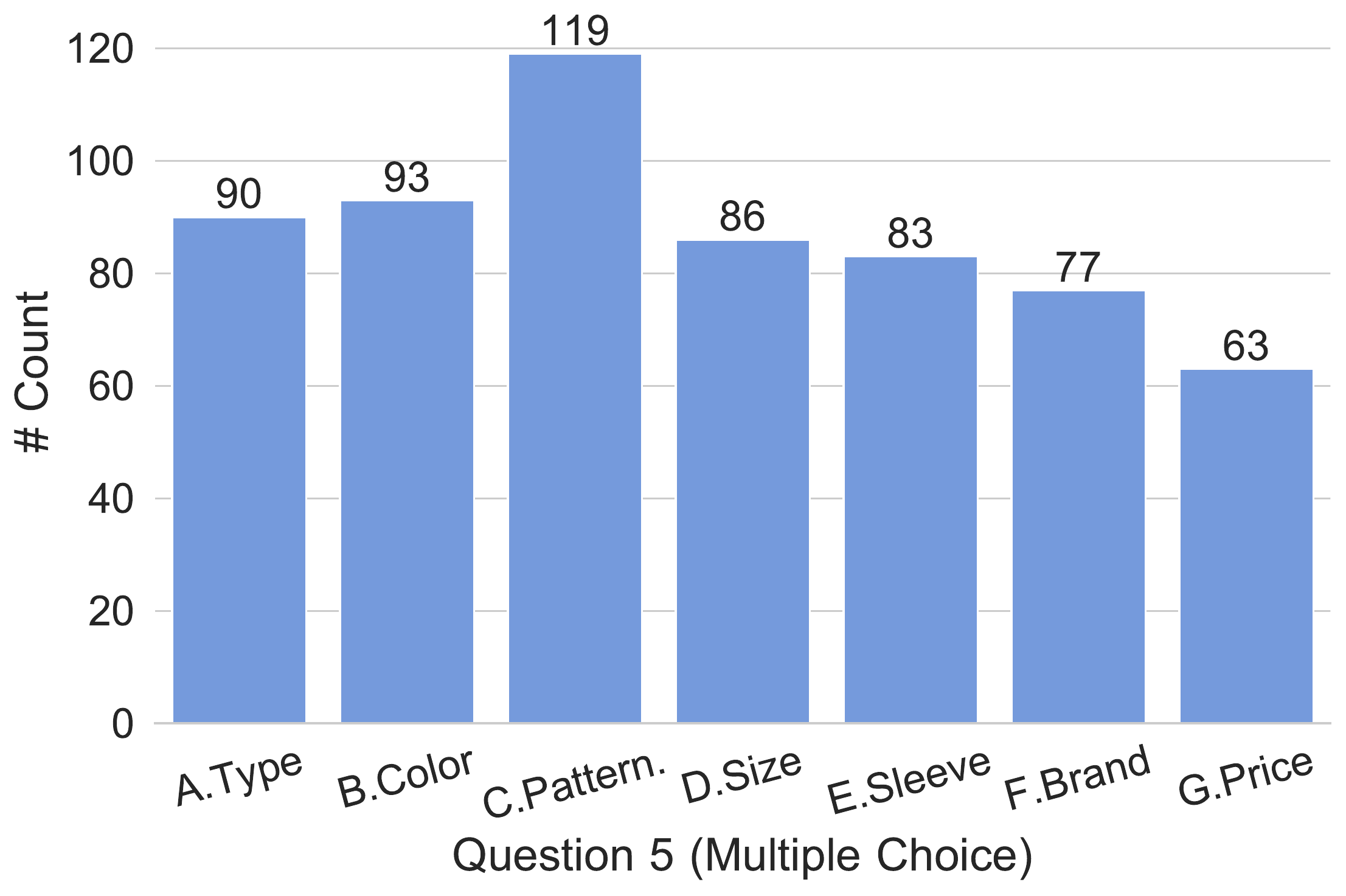}
    \caption{The bar chart of Question 5 results. From the choice distribution, we can see that \textbf{Pattern} plays the most important role in successful recommendation while \textbf{Price} is relatively less essential.}
    \label{fig:question_5}
    \vspace{-0.4cm}
\end{figure}

\begin{figure}[htp]
    \centering
    \includegraphics[width=0.7\linewidth]{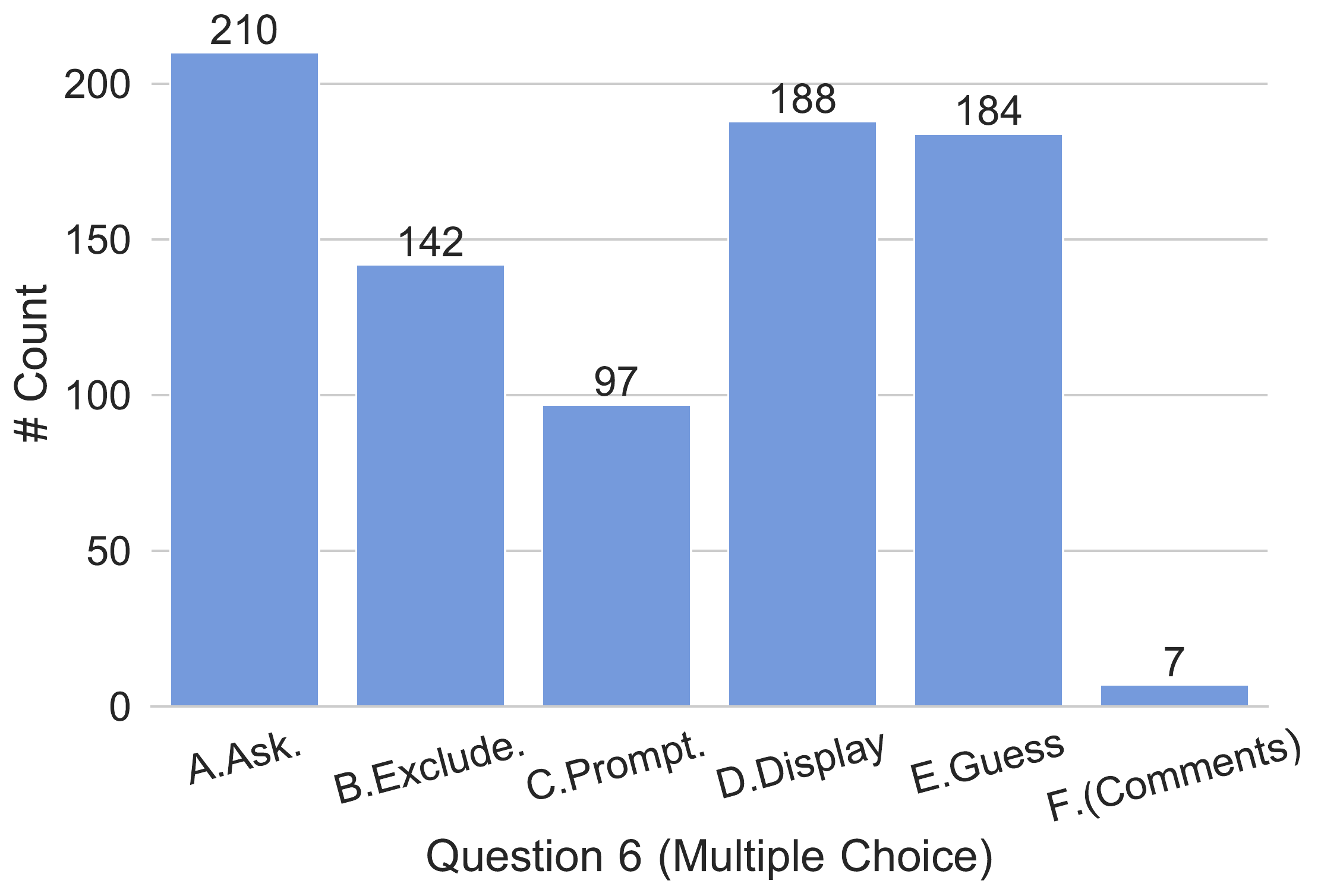}
    \caption{The bar chart of Question 6 results. From the choice distribution, we can see that salespersons are prone to ask preference, display candidate attribute values and guess one concrete attribute value. Sometimes, they also try to exclude customer unlikeness or actively prompt.}
    \label{fig:question_6}
    \vspace{-0.4cm}
\end{figure}

\begin{figure}[htp]
    \centering
    \includegraphics[width=0.7\linewidth]{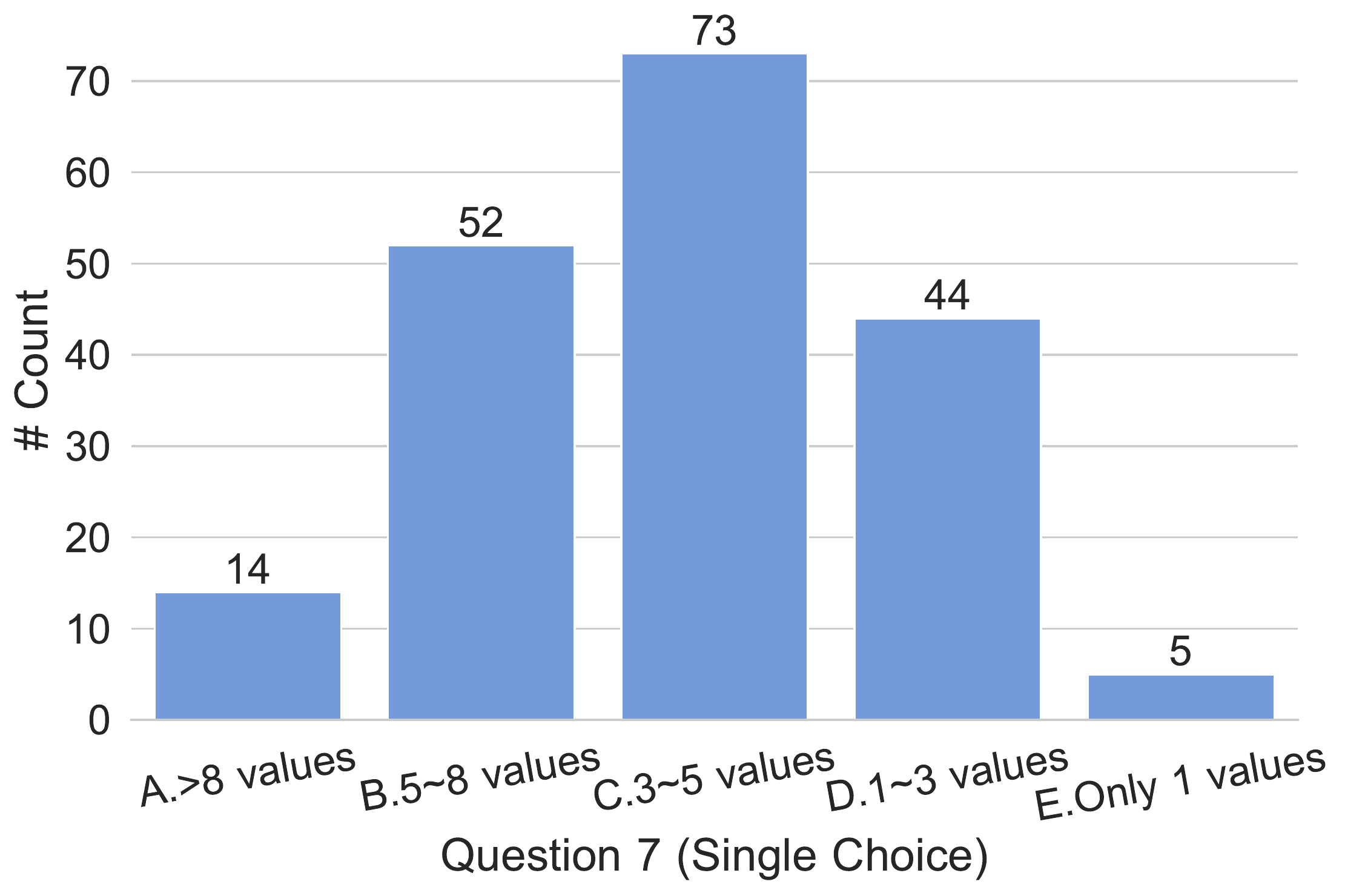}
    \caption{The bar chart of Question 7 results. Only 188 sales experts who choose Choice D in Question 6 participate in answering this question. We can find that most sales experts prefer to display all candidate attribute values when there only exists 3\textasciitilde5 values in the candidate set.}
    \label{fig:question_7}
    \vspace{-0.4cm}
\end{figure}

\begin{figure}[htp]
    \centering
    \includegraphics[width=0.5\linewidth]{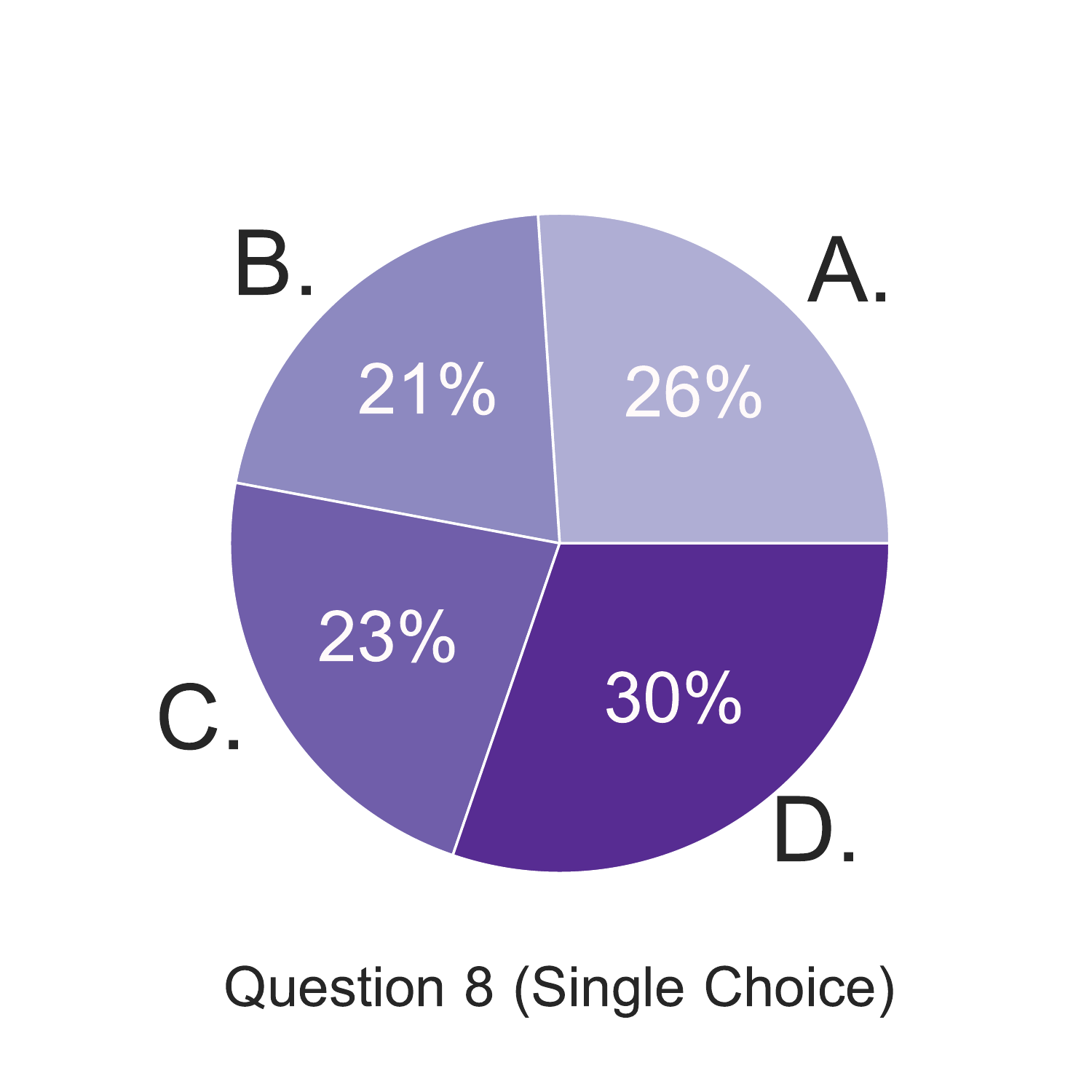}
    \caption{The pie chart of Question 8 results. Most sales experts are prone to learn about some aspects of customer preferences in detail as \textbf{Choice B}, \textbf{Choice C} and \textbf{Choice D} show. Only 21\% of sales experts just roughly learn about customer preferences before recommending items.}
    \label{fig:question_8}
    \vspace{-0.4cm}
\end{figure}

\begin{figure}[htp]
    \centering
    \includegraphics[width=0.5\linewidth]{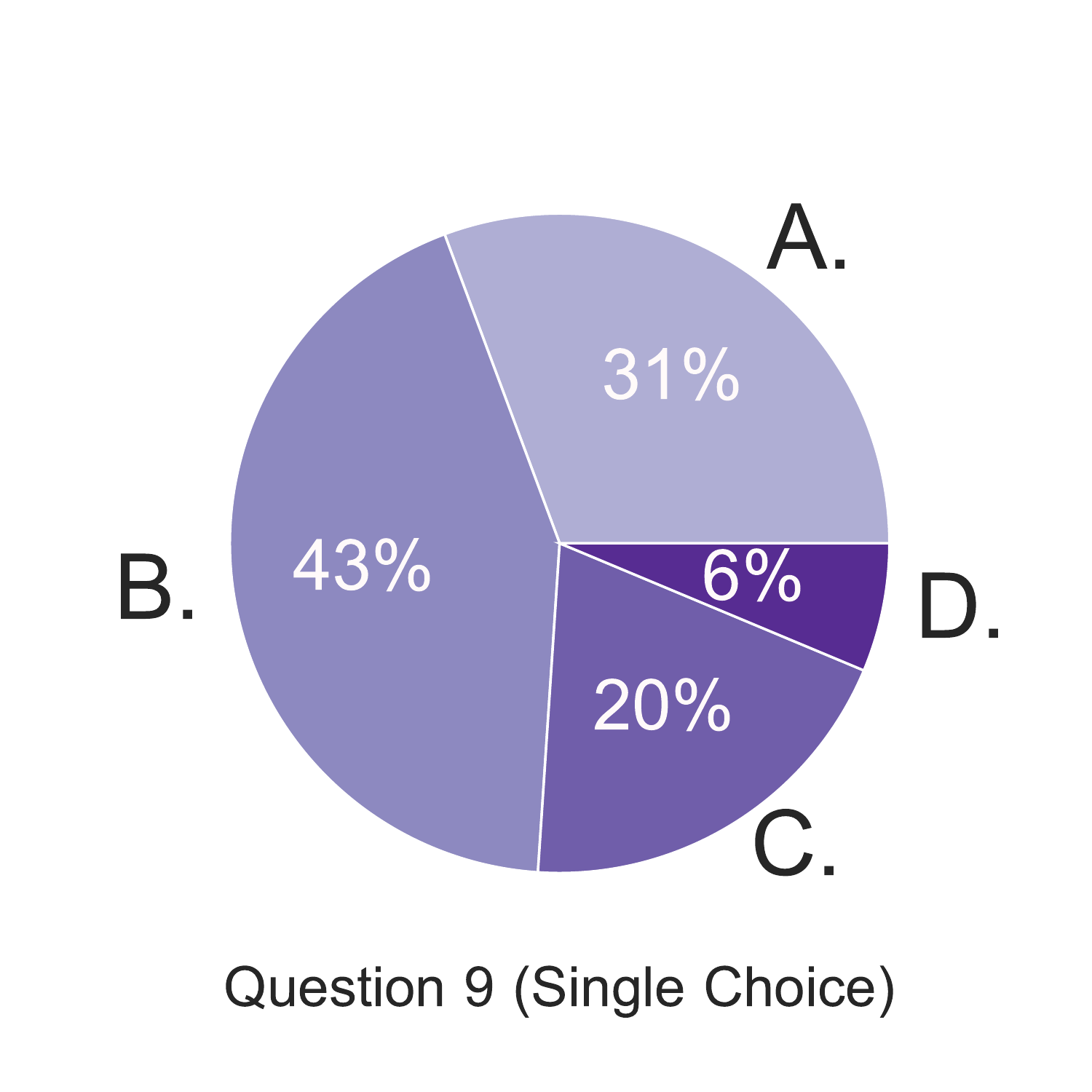}
    \caption{The pie chart of Question 9 results. We can find that most sales experts (\textbf{Choice A} and \textbf{Choice B}) narrow candidate items range by referring one specific region in the store scene after they elicit several aspects of customer preference.}
    \label{fig:question_9}
    \vspace{-0.3cm}
\end{figure}

\begin{figure}[htp]
    \centering
    \includegraphics[width=0.7\linewidth]{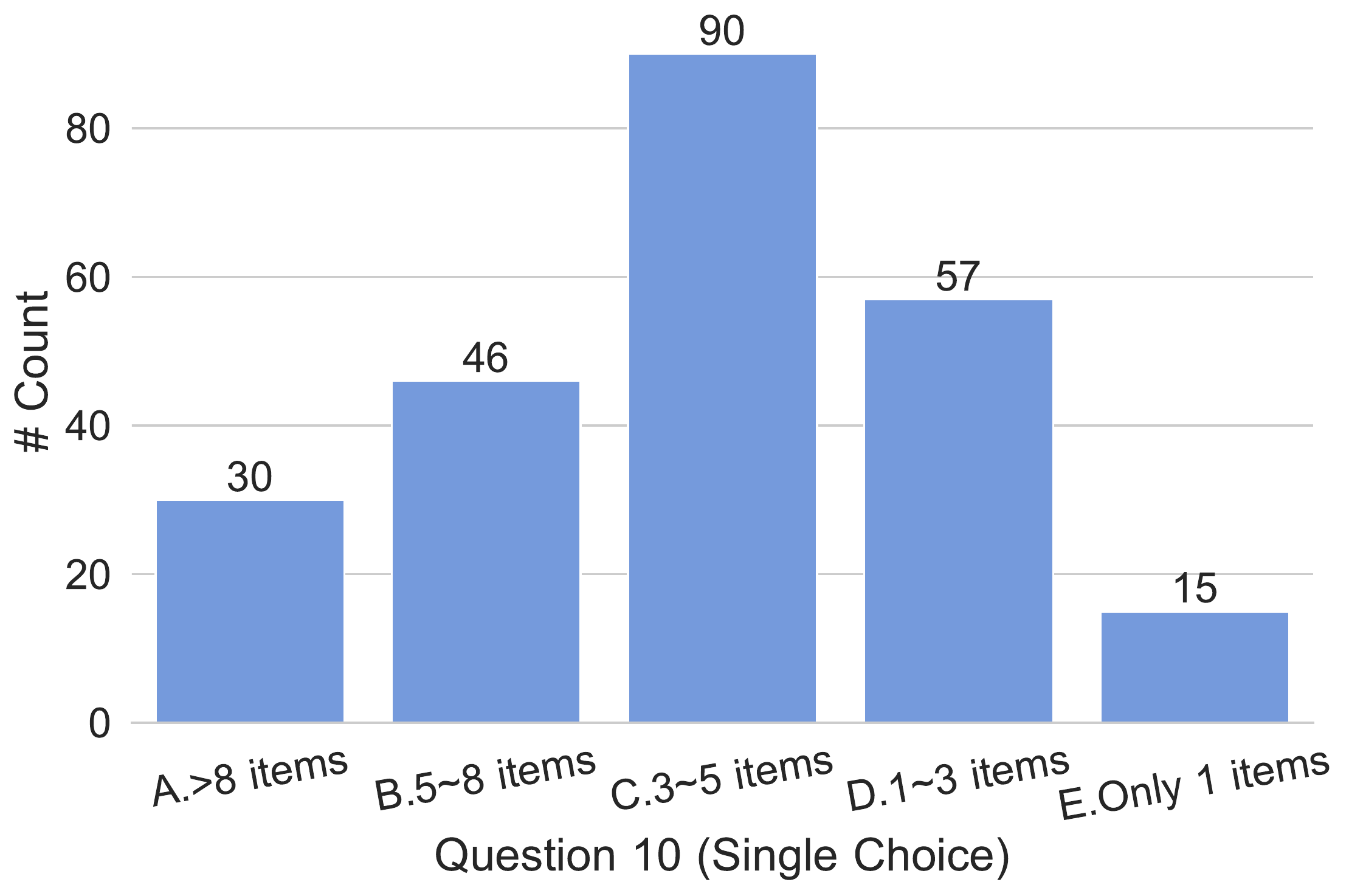}
    \caption{The bar chart of Question 10 results. We can find that the distribution mainly falls into \textbf{Choice C} and \textbf{Choice D}, which indicates that sales experts prefer to recommend items when there are about 1\textasciitilde5 items left in the candidate set.}
    \label{fig:question_10}
    \vspace{-0.4cm}
\end{figure}

\begin{figure}[htp]
    \centering
    \includegraphics[width=0.7\linewidth]{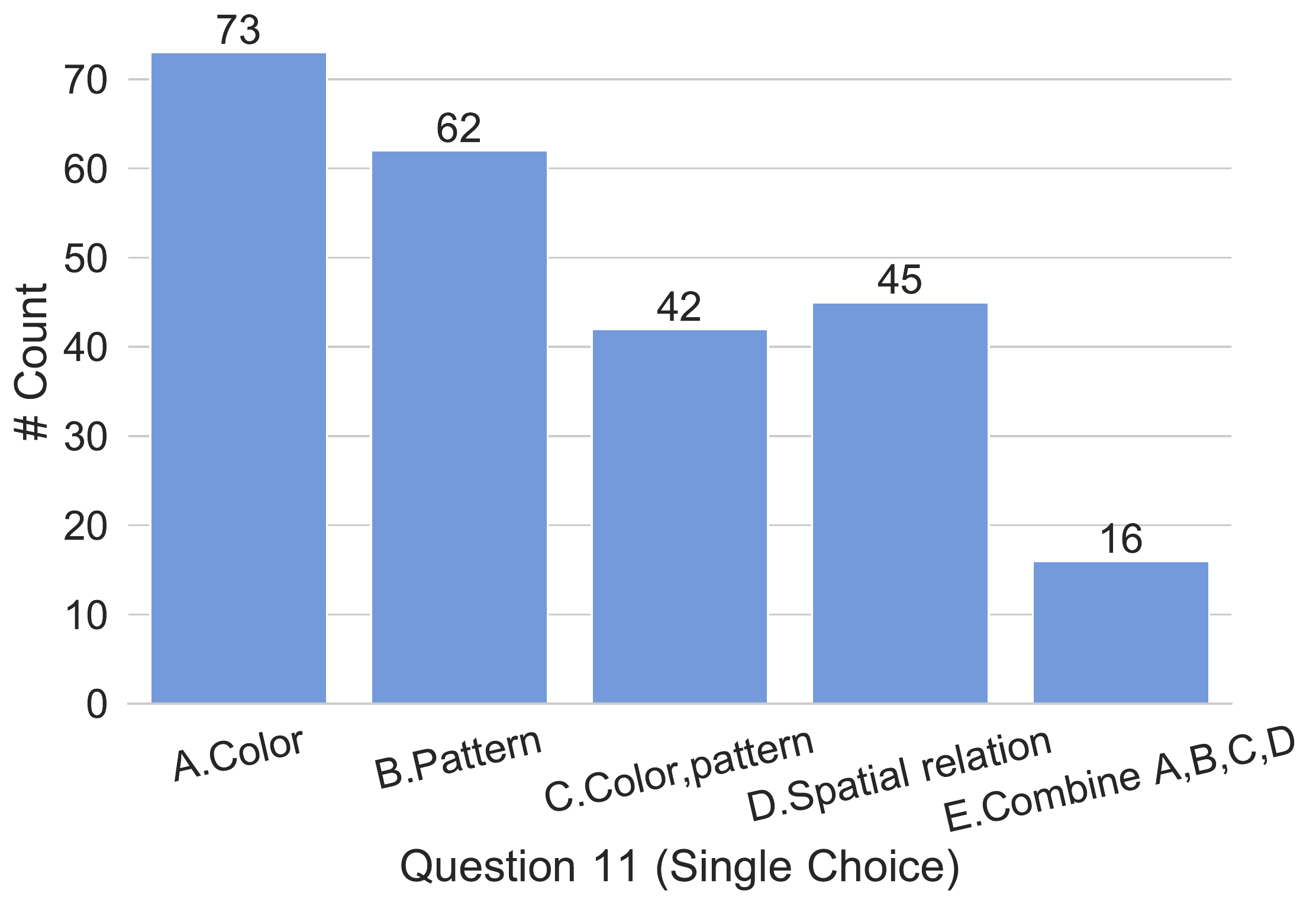}
    \caption{The bar chart of Question 11 results. Most sales experts are prone to refer scene items solely by \textbf{Color}, \textbf{Pattern} or \textbf{Spatial Relation} instead of their combinations. Clear and simple referring expression is more consistent with real-life shopping conversation.}
    \label{fig:question_11}
    \vspace{-0.4cm}
\end{figure}

\begin{figure}[htp]
    \centering
    \includegraphics[width=0.7\linewidth]{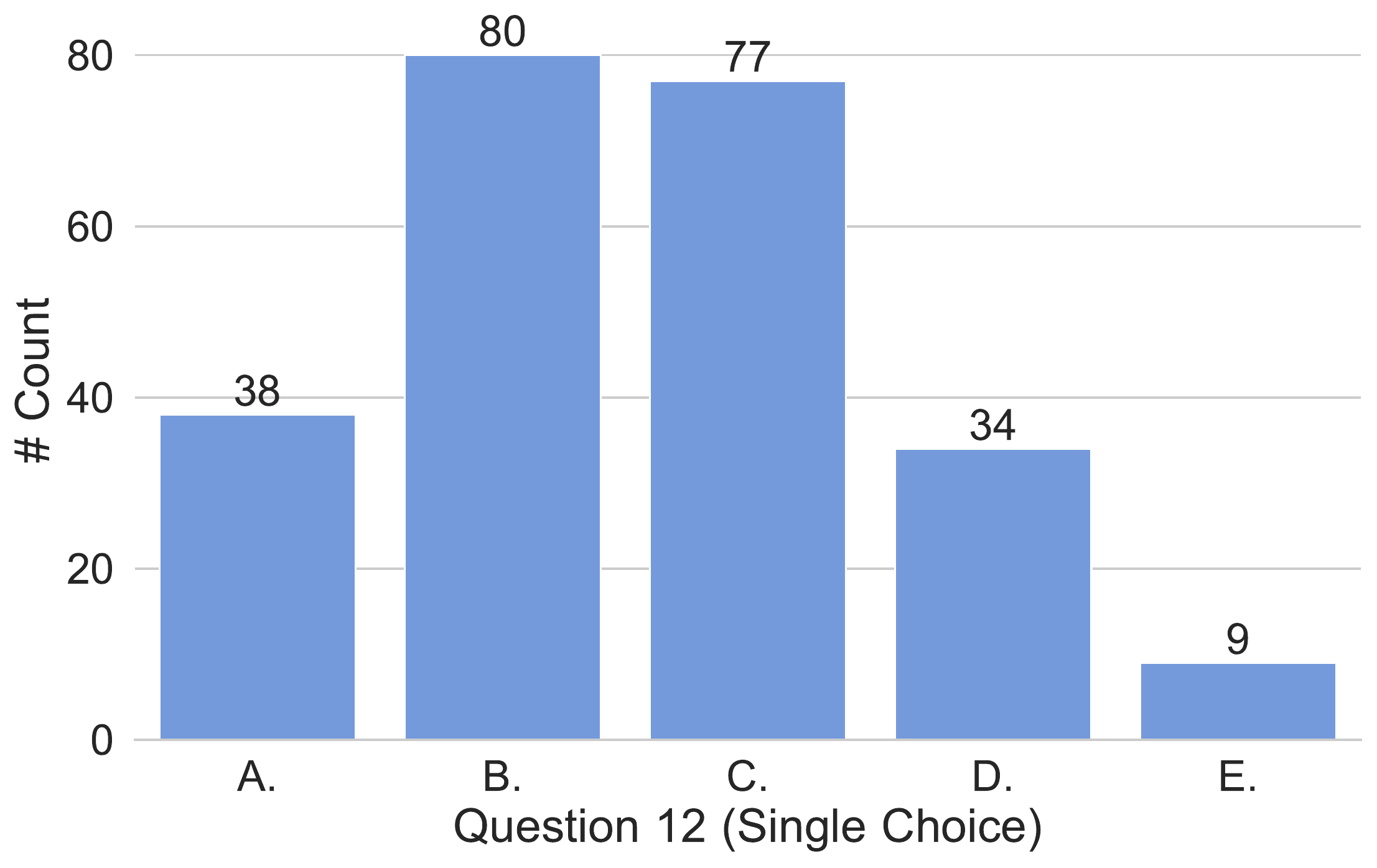}
    \caption{The bar chart of Question 12 results. We find that sales experts pay more attention to reality and diversity of recommendation strategies (\textbf{Choice B}) and customer shopping experience (\textbf{Choice C}) rather than recommendation efficiency (\textbf{Choice A}).}
    \label{fig:question_12}
    \vspace{-0.4cm}
\end{figure}

For clothes recommendation questionnaire, we totally receive 765 complete survey results. To guarantee the reliability of servery, we exclude questionnaires from salesperson with less than three years work experience and remain 238 results for statistics. In the following, we visualize survey results in Fig.~\ref{fig:question_1} to Fig.~\ref{fig:question_12} by bar charts and pie charts. 

\begin{figure}[htp]
    \centering
    \includegraphics[width=0.7\linewidth]{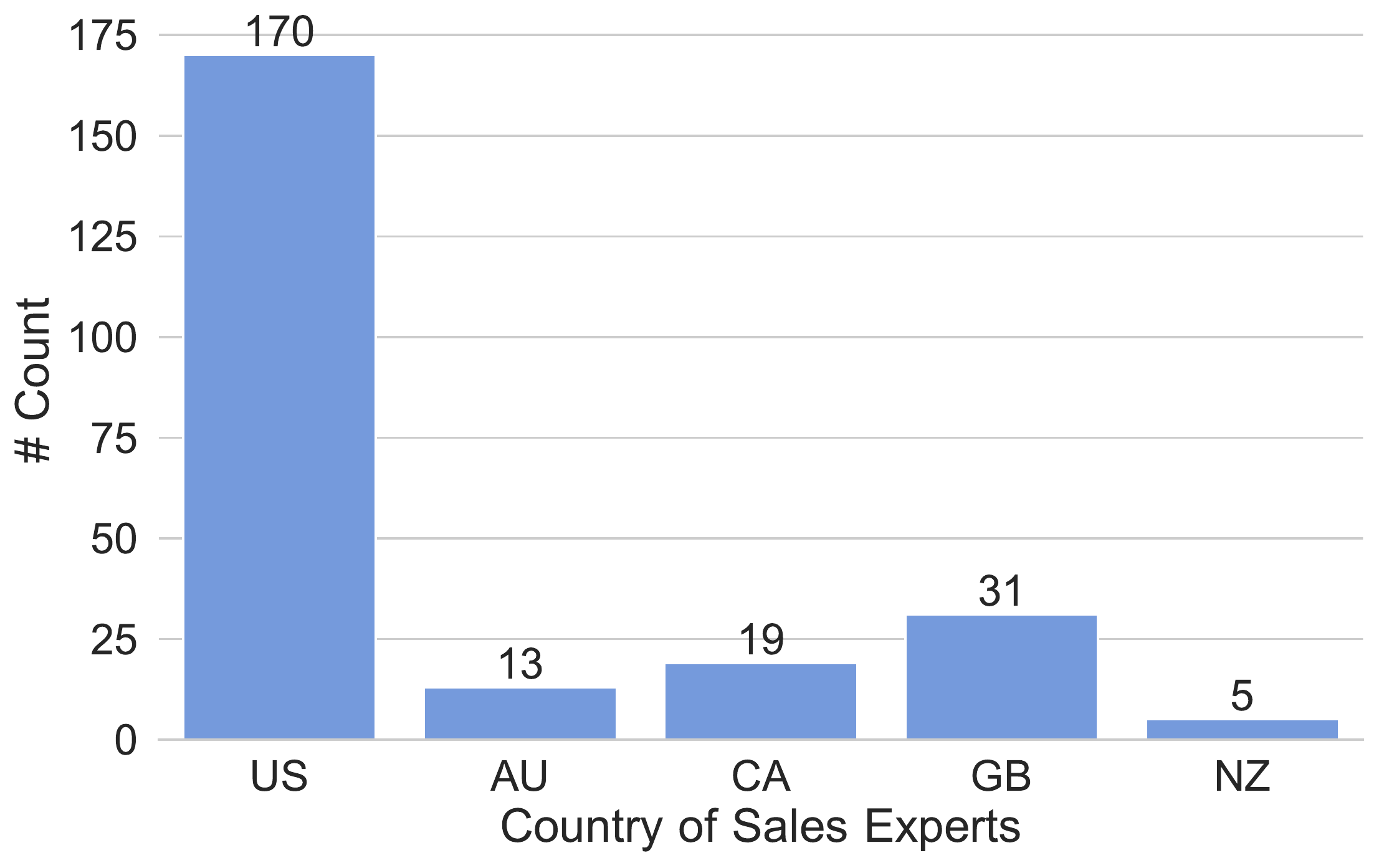}
    \caption{The bar chart displays the demographic and geographic characteristics of the 238 surveyed sales experts. Most experts come from United States (US) while other experts come from United Kingdom (GB), Australia (AU), Canada (CA) and New Zealand (NZ).}
    \label{fig:population}
    \vspace{-0.4cm}
\end{figure}

\subsubsection{Manual Paraphrase} \label{sssec:paraphrase}
We release "Dialog Writing with Subjective Preference" task on on Amazon Mechanical Turk (AMT) platform to hire workers to write subjective preferences according to attribute categorization concepts and paraphrase dialog flows to natural language utterances with subjective preferences. To guarantee the quality, we require answers have greater than 90\% HIT approval rate. For payment, we pay for \$1 for every carefully paraphrased dialog, which is competitive with similar tasks in the same period. We display the instruction and scene snapshot for this task in the following. \\

\centerline{\textit{\textbf{Dialog Writing with Subjective Preference}}}
\textit{This task is to collect multimodal recommendation dialogs between salesperson and customer with subjective preferences. You need to write subjective preferences based on given attribute categorization concepts by visual perception or commonsense firstly. For example, \textbf{"warm color"} can be written to "color of passion", "color of happiness" and "color for welcome ceremony" while \textbf{"lively pattern"} can be written to "vibrant pattern", "pattern closer to nature" and "pattern that is popular among conservationists". Then, you need to paraphrase provided dialog flow to dialog with written subjective preferences. We introduce every kind of dialog act in the following.} 

\textit{
\begin{itemize}
    \item[$\bullet$] \textbf{\textit{Ask Preference (Salesperson)}} refers to asking customer preference about one attribute type. 
    \item[$\bullet$] \textbf{\textit{Answer Preference (Customer)}} refers to answering subjective preference about asked attribute type. 
    \item[$\bullet$] \textbf{\textit{Exclude Preference (Salesperson)}} is asking customer what he unlike. 
    \item[$\bullet$] \textbf{\textit{Negate Preference (Customer)}} is answering unlike attribute by subjective preference. 
    \item[$\bullet$] \textbf{\textit{Prompt Preference (Salesperson)}} is actively providing subjective "preference" to inspire customer. 
    \item[$\bullet$] \textbf{\textit{Respond Prompt (Customer)}} is responding to salesperson's prompt.
    \item[$\bullet$] \textbf{\textit{Guess Attribute Value (Salesperson)}} is predicting one concrete value from candidate attribute value.
    \item[$\bullet$] \textbf{\textit{Revise Attribute Value (Salesperson)}} is revising prediction of concrete attribute value following customer's feedback.
    \item[$\bullet$] \textbf{\textit{Respond Attribute Value (Customer)}} is responding to attribute value guessed or revised by salerperson.
    \item[$\bullet$] \textbf{\textit{Display Candidate Values (Salesperson)}} refers to listing all candidate attribute values for customer to choose.
    \item[$\bullet$] \textbf{\textit{Choose Value (Customer)}} refers to choosing target value from listed values.
    \item[$\bullet$] \textbf{\textit{Refer Region (Salesperson)}} is an act that salesperson points out one region in the store to ask customer whether the region contains item he wants.
    \item[$\bullet$] \textbf{\textit{Judge Region (Customer)}} is judging whether the referred region contains the item that customer is looking for.
    \item[$\bullet$] \textbf{\textit{Recommend Item (Salesperson)}} is trying to recommend items from candidate item set based on multimodal context. 
    \item[$\bullet$] \textbf{\textit{Respond Recommend (Customer)}} is responding to item recommendation.
\end{itemize}
} 

\textit{Please paraphrase dialog act according to above information. The written should keep the similar meaning as original dialog act and reserve the corresponding attribute type or subjective preference following given slot. For example, "Answer\_Preference:\{Color:warm color\}" can be written to "I am looking for clothes with color that makes me feel happy." while "Negate\_Preference:\{Pattern:dazzling pattern\}" can be written to "I am interested in eyes-catching pattern." You are welcome to add visual descriptions and spatial relations with background items to refer commodity items (orange bounding boxes in the scene snapshot annotate all background items). You are encouraged to add polite expressions and modal particles into utterances. Note that the hit will be rejected if the utterances contain any racism, sexism and privacy information. The annotated dialogs may be presented at scientific meetings or published in scientific journals for academic research. If you are fully aware of and agree with above information, you are welcome to accept the task.}

\begin{figure}[htp]
    \centering
    \includegraphics[width=1\linewidth]{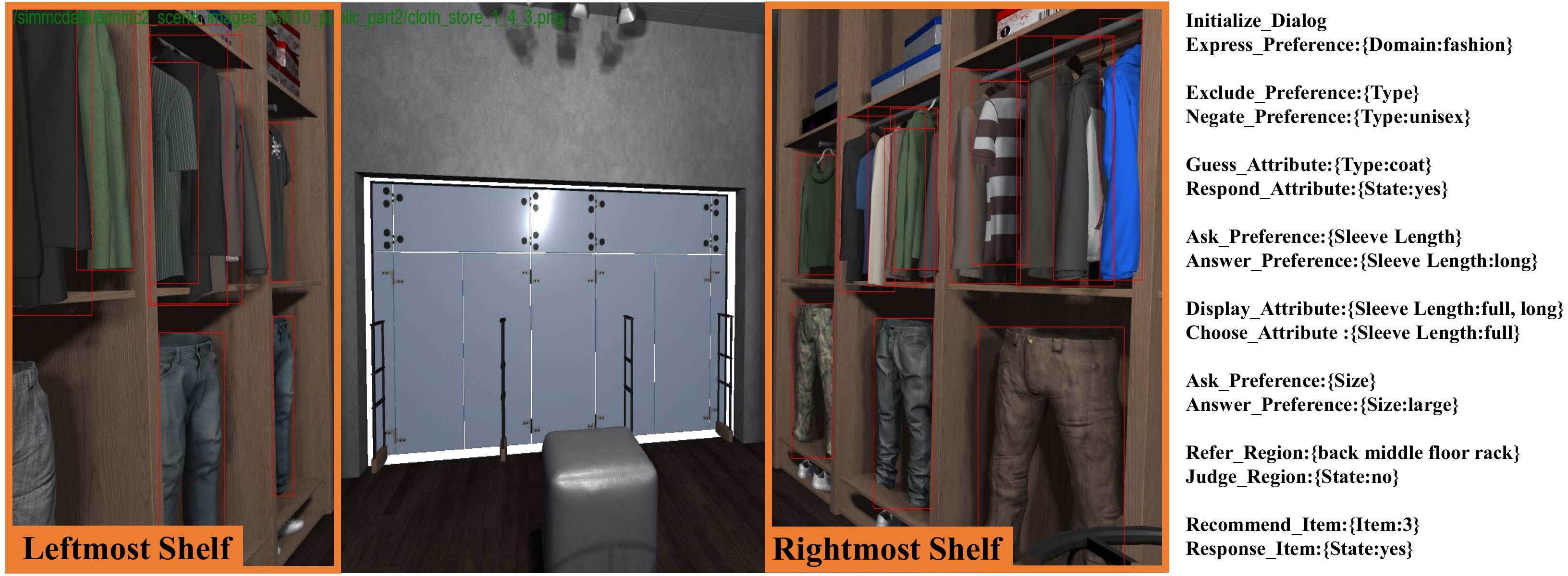}
    \caption{Scene snapshot of manual paraphrase. Red bounding boxes are for clothes items while orange boxes are for background items.}
    \label{fig:paraphrase}
    \vspace{-0.3cm}
\end{figure}

\subsection{Case Study for MRA}

\begin{figure}[htp]
    \centering
    \includegraphics[width=1\linewidth]{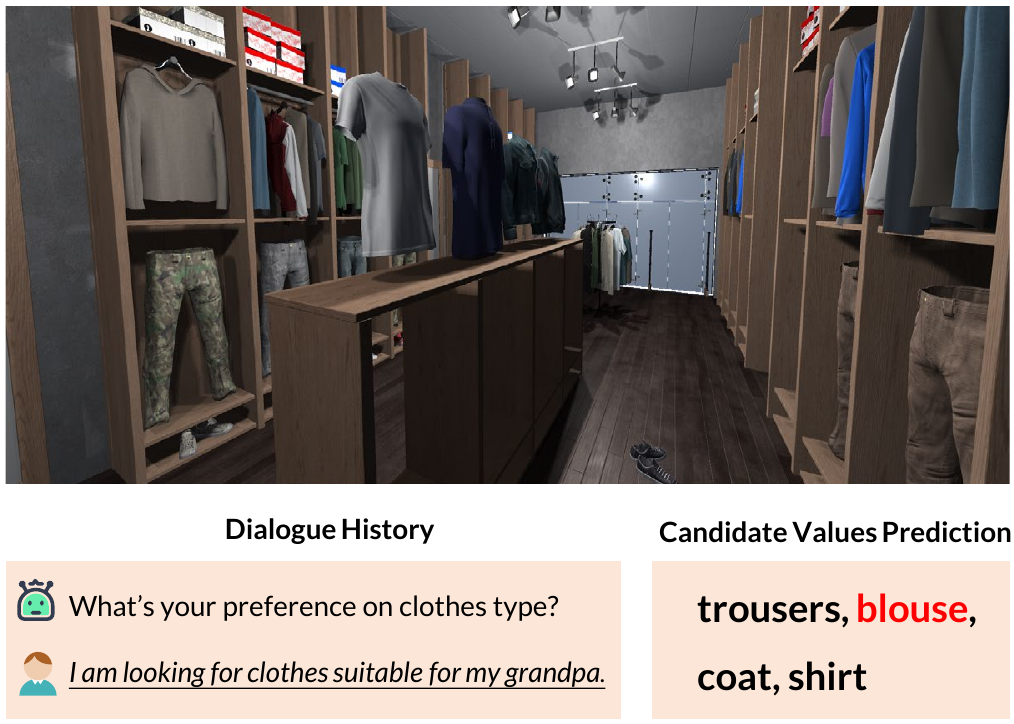}
    \caption{Case study for MRA.}
    \label{fig:case_study}
    \vspace{-0.3cm}
\end{figure}

\begin{figure*}[htp]
    \centering
    \includegraphics[width=1\linewidth]{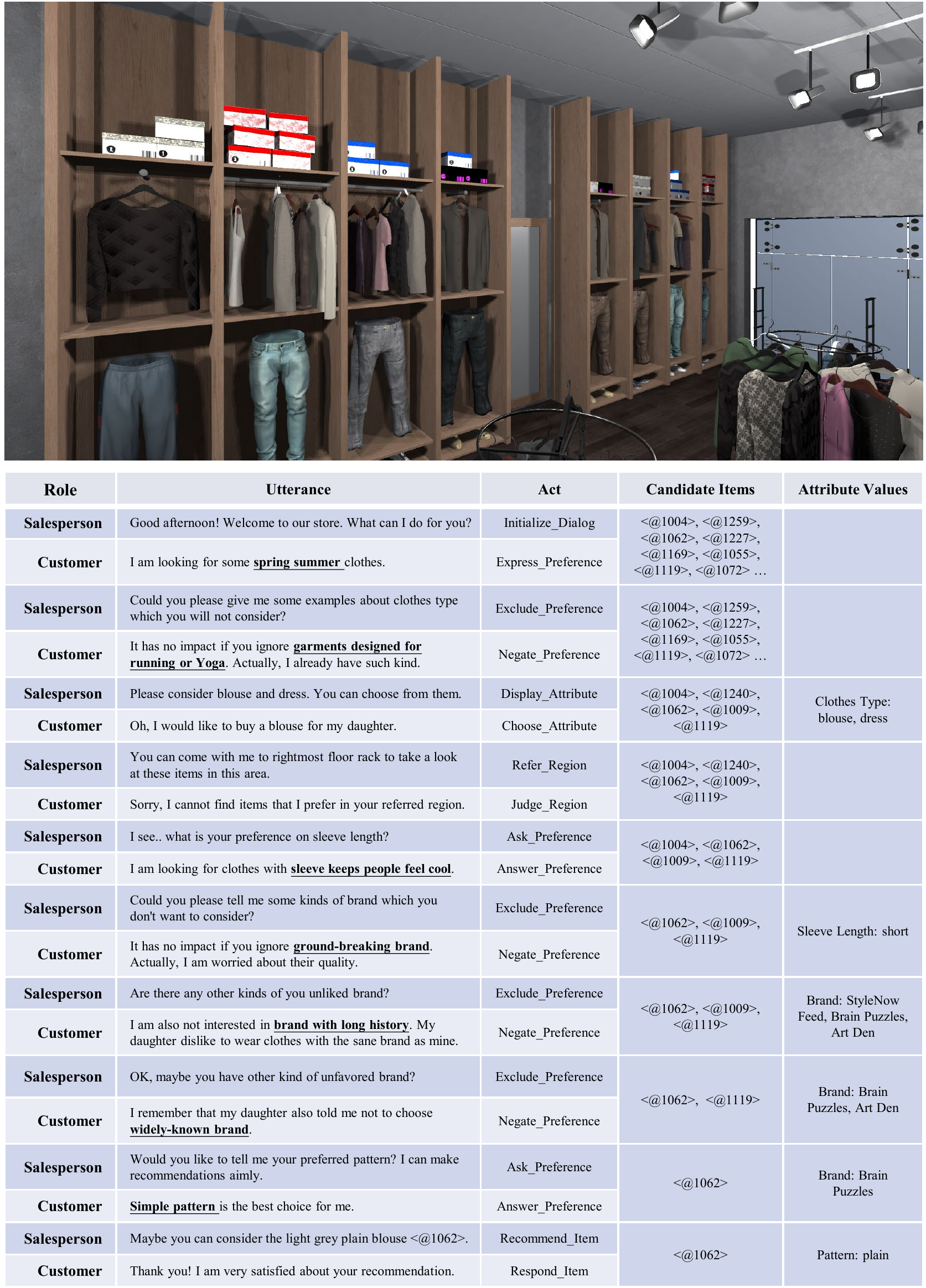}
    \caption{Example of one complete dialog section in SURE dataset. Candidate Items and candidate attribute values in the last two columns are derived from previous rounds and situated scene. We can observe that salesperson adopt acts including \textbf{\textit{"Ask\_Preference"}}, \textbf{\textit{"Exclude\_Preference"}}, \textbf{\textit{"Display\_Attribute"}} to clarify customer subjective preferences highlighted by underline and narrow candidate range of items by \textbf{\textit{"Refer\_Region"}}.}
    \label{fig:dialog_case}
\end{figure*}

\begin{figure*}[t]
    \centering
    \includegraphics[width=1\linewidth]{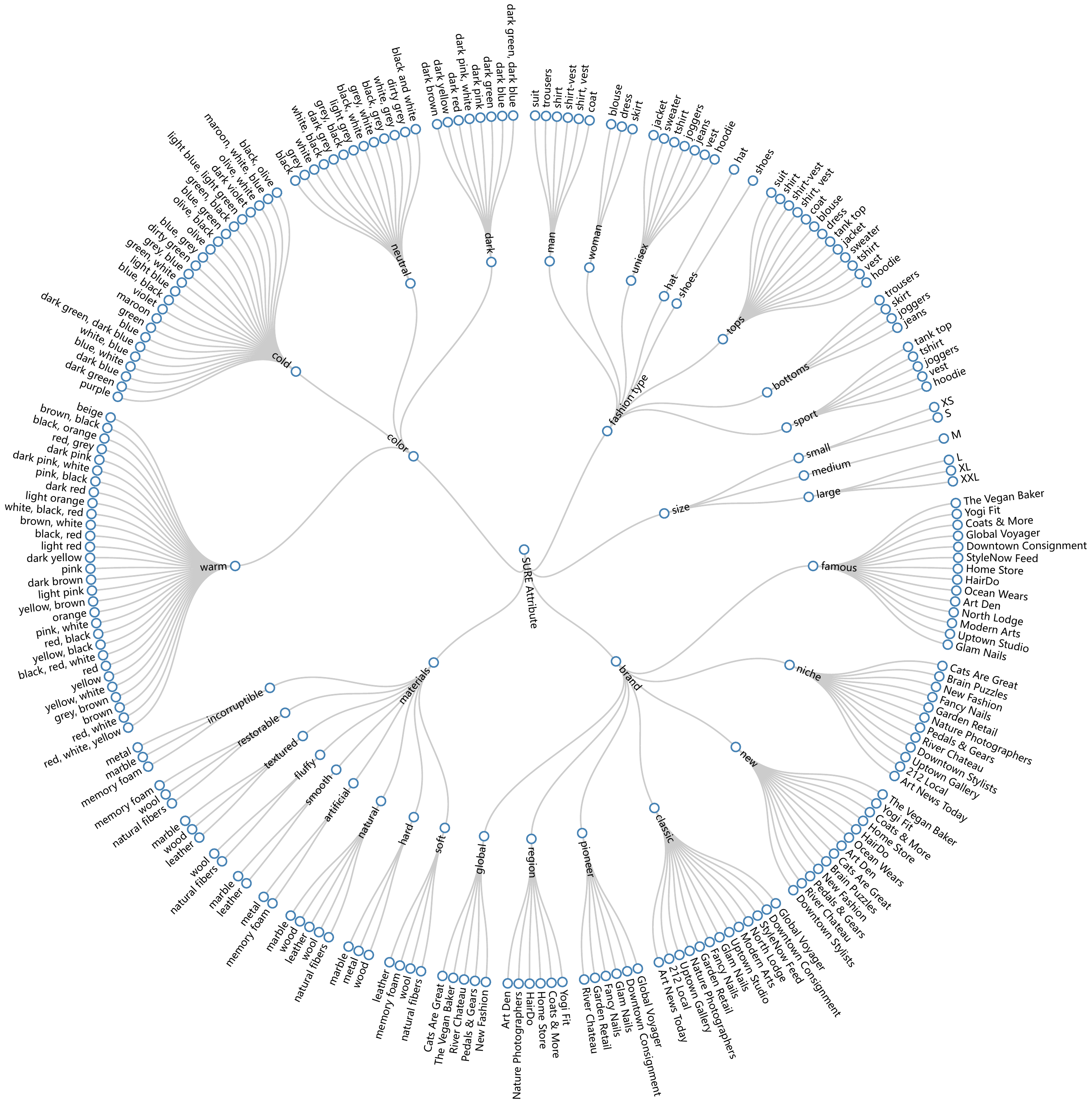}
    \caption{Some attribute categorization concepts and corresponding concrete attribute values in SURE dataset.}
    \label{fig:redial_tree}
\end{figure*}

\end{document}